%% file: main.tex
\documentclass[11pt,a4paper]{article}
\pdfoutput=1
\usepackage{jheppub}
\usepackage{microtype}
\usepackage{dcolumn}
\usepackage{booktabs}

\def\bfseries{\fontseries \bfdefault \selectfont \boldmath}

\input{src/Commands.tex}

\begin{document}

\title{First proof of topological signature in the high pressure xenon gas TPC with electroluminescence amplification for the NEXT experiment}

\input{src/Authors}

\abstract{The NEXT experiment aims to observe the neutrinoless double
beta decay of \XE\ in a high-pressure xenon gas TPC using
electroluminescence (EL) to amplify the signal from
ionization. One of the main advantages of this technology is the
possibility to reconstruct the topology of events with energies close
to \Qbb. This paper presents the first demonstration that the
topology provides extra handles to reject background events using data
obtained with the NEXT-DEMO prototype.

Single electrons resulting from the interactions of \NA\ 1275~keV gammas and electron-positron pairs produced by conversions of gammas from the \THO\ decay chain were used to represent the background and the signal in a double beta decay. These data were used to develop algorithms for the reconstruction of tracks and the identification of the energy deposited at the end-points, providing an extra background rejection factor of $24.3 \pm 1.4$ (stat.)\%, while maintaining an efficiency of  $66.7 \pm 1.$ \% for signal events. 
}
\maketitle

\clearpage

\section{Introduction}
\label{intro}
\input{src/Intro.tex}

\section{Topological signature in NEXT}
\label{Toposig}
\input{src/TopoSig.tex}

\section{Analysis}
\label{Analysis}
\input{src/Analysis}

\subsection{Data set}
\label{Data}
\input{src/Data.tex}
\subsection{Event selection}
\label{Selection}
\input{src/Selection.tex}
\section{Results}
\label{Results}
\input{src/Results.tex}

\section{Conclusions}
\label{Conclus}
\input{src/Conclus.tex}

\acknowledgments
This work was supported by the following agencies and institutions: the \emph{European Research Council} (ERC) under the Advanced Grant 339787-NEXT; the \emph{Ministerio de Econom\'ia y Competitividad} of Spain under grants CONSOLIDER-Ingenio 2010 CSD2008-0037 (CUP), FPA2009-13697-C04 and FIS2012-37947-C04; the Portuguese FCT and FEDER through the program COMPETE, project PTDC/FIS/103860/2008; and the Fermi National Accelerator Laboratory under U.S. Department of Energy Contract No. DE-AC02-07CH11359.

\bibliographystyle{JHEP}
\bibliography{biblio}

\end{document}

%% file: src/Commands.tex
\newcommand{\bbonu}{\ensuremath{\beta\beta0\nu}}
\newcommand{\bbtnu}{\ensuremath{\beta\beta2\nu}}
\newcommand{\Qbb}{\ensuremath{Q_{\beta\beta}}}

\newcommand{\NA}{\ensuremath{^{22}}Na}

\newcommand{\XE}{\ensuremath{{}^{136}\rm Xe}}
\newcommand{\TL}{\ensuremath{{}^{208}\rm{Tl}}}
\newcommand{\BI}{\ensuremath{^{214}}Bi}

\newcommand{\UDTO}{\ensuremath{{}^{238}\rm U}}

\newcommand{\THO}{\ensuremath{{}^{228}{\rm Th}}}

%% file: src/Authors.tex
\collaboration{The NEXT Collaboration}

\author[a,1]{P.~Ferrario,\note{Corresponding author}}
\author[a]{A.~Laing,}
\author[a] {N.~L\'opez-March,}
\author[a,2]{J.J.~G\'omez-Cadenas,\note{Spokesperson}}

\author[a] {V.~\'Alvarez,}
\author[g] {C.D.R.~Azevedo,}
\author[b] {F.I.G.~Borges,}
\author[a] {S.~C\'arcel,}
\author[c] {S.~Cebri\'an,}
\author[a] {A.~Cervera,}
\author[b] {C.A.N.~Conde,}
\author[c] {T.~Dafni,}
\author[a] {J.~D\'iaz,}
\author[o]{ M.~Diesburg,}
\author[e] {R.~Esteve,}
\author[b] {L.M.P.~Fernandes,}
\author[g]{ A.L.~Ferreira,}
\author[b]{ E.D.C.~Freitas,}
\author[d]{V.M.~Gehman,}
\author[d]{ A.~Goldschmidt,}
\author[p]{ D.~Gonz\'alez-D\'iaz,}
\author[h]{ R.M.~Guti\'errez,}
\author[i]{ J.~Hauptman,}
\author[b]{C.A.O.~Henriques,}
\author[j]{ J.A.~Hernando Morata,}
\author[c]{ I.G.~Irastorza,}
\author[k]{ L.~Labarga,}
\author[o]{ P.~Lebrun,}
\author[a]{ I.~Liubarsky,}
\author[a]{ D.~Lorca,}
\author[h]{ M.~Losada,}
\author[c]{ G.~Luz\'on,}
\author[e]{ A.~Mar\'i,}
\author[a]{ J.~Mart\'in-Albo,}
\author[j]{ G.~Mart\'inez-Lema,}
\author[a]{ A.~Mart\'inez,}
\author[d]{ T.~Miller,}
\author[a]{ F.~Monrabal,}
\author[a]{ M.~Monserrate,}
\author[b]{ C.M.B.~Monteiro,}
\author[e]{ F.J.~Mora,}
\author[g]{ L.M.~Moutinho,}
\author[a]{ J.~Mu\~noz~Vidal,}
\author[a]{ M.~Nebot-Guinot,}
\author[a]{ P.~Novella,}
\author[d]{ D.~Nygren,}
\author[o]{ A.~Para,}
\author[k]{ J.~P\'erez,}
\author[l]{ J.L.~P\'erez~Aparicio,}
\author[a]{ M.~Querol,}
\author[a]{ J.~Renner,}
\author[m]{ L.~Ripoll,}
\author[a]{ J.~Rodr\'iguez,}
\author[b]{ F.P.~Santos,}
\author[b]{ J.M.F.~dos~Santos,}
\author[a]{ L.~Serra,}
\author[d]{ D.~Shuman,}
\author[a]{ A.~Sim\'on,}
\author[n]{ C.~Sofka,}
\author[a]{ M.~Sorel,}
\author[d]{ J.F.~Toledo,}
\author[m]{ J.~Torrent,}
\author[f]{ Z.~Tsamalaidze,}
\author[g]{ J.F.C.A.~Veloso,}
\author[c]{ J.A.~Villar,}
\author[n]{ R.~Webb,}
\author[n,3]{ J.T.~White\note{deceased.},}
\author[a]{ N.~Yahlali,}
\author[h] {H.~Yepes-Ram\'irez}

\emailAdd{paola.ferrario@ific.uv.es}

\affiliation[a]{
Instituto de F\'isica Corpuscular (IFIC), CSIC \& Universitat de Val\`encia\\
Calle Catedr\'atico Jos\'e Beltr\'an, 2, 46980 Paterna, Valencia, Spain}
\affiliation[b]{
Departamento de Fisica, Universidade de Coimbra\\
Rua Larga, 3004-516 Coimbra, Portugal}
\affiliation[c]{
Lab.\ de F\'isica Nuclear y Astropart\'iculas, Universidad de Zaragoza\\ 
Calle Pedro Cerbuna, 12, 50009 Zaragoza, Spain}
\affiliation[d] {
Lawrence Berkeley National Laboratory (LBNL)\\
1 Cyclotron Road, Berkeley, California 94720, USA}
\affiliation[e] {
Instituto de Instrumentaci\'on para Imagen Molecular (I3M), Universitat Polit\`ecnica de Val\`encia\\ 
Camino de Vera, s/n, Edificio 8B, 46022 Valencia, Spain}
\affiliation[f] {
Joint Institute for Nuclear Research (JINR)\\
Joliot-Curie 6, 141980 Dubna, Russia}
\affiliation[g] {
Institute of Nanostructures, Nanomodelling and Nanofabrication (i3N), Universidade de Aveiro\\
Campus de Santiago, 3810-193 Aveiro, Portugal}
\affiliation[h] {
Centro de Investigaciones, Universidad Antonio Nari\~no\\ 
Carretera 3 este No.\ 47A-15, Bogot\'a, Colombia}
\affiliation[i] {
Department of Physics and Astronomy, Iowa State University\\
12 Physics Hall, Ames, Iowa 50011-3160, USA}
\affiliation[j] {
Instituto Gallego de F\'isica de Altas Energ\'ias (IGFAE), Univ.\ de Santiago de Compostela\\
Campus sur, R\'ua Xos\'e Mar\'ia Su\'arez N\'u\~nez, s/n, 15782 Santiago de Compostela, Spain}
\affiliation[k] {
Departamento de F\'isica Te\'orica, Universidad Aut\'onoma de Madrid\\
Ciudad Universitaria de Cantoblanco, 28049 Madrid, Spain}
\affiliation[l] {
Dpto.\ de Mec\'anica de Medios Continuos y Teor\'ia de Estructuras, Univ.\ Polit\`ecnica de Val\`encia\\
Camino de Vera, s/n, 46071 Valencia, Spain}
\affiliation[m] {
Escola Polit\`ecnica Superior, Universitat de Girona\\
Av.~Montilivi, s/n, 17071 Girona, Spain}
\affiliation[n] {
Department of Physics and Astronomy, Texas A\&M University\\
College Station, Texas 77843-4242, USA}
\affiliation[o] {
Fermi National Accelerator Laboratory\\
Batavia, Illinois 60510, USA}
\affiliation[p] {
CERN, European Organization for Nuclear Research\\
1211 Geneva, Switzerland}

%% file: src/Intro.tex
The observation of neutrino oscillations by several experiments proves that neutrinos have mass, as described, for instance, in Ref. \cite{GonzalezGarcia:2007ib}. This, however, motivates a number of questions about the nature of neutrinos and their relation to the other fermions: How can this mass be accommodated in the Standard Model? What is the absolute scale of neutrino mass? Does CP violation take place in neutrino oscillations? Could the neutrino be its own antiparticle?

All other fermions in the Standard Model (quarks and charged leptons) have their mass described by a Dirac mass term alone. Neutrino mass could also be generated by the same mechanism, however, the neutrino, being chargeless, can also have a Majorana mass term. Majorana neutrinos would be indistinguishable from their antiparticle and may lead to processes violating lepton number. Such processes may be responsible for the generation of the observed matter-antimatter asymmetry of the Universe via leptogenesis \cite{Fukugita:1986hr}. The Majorana mass of the neutrino also provides an elegant explanation of the smallness of neutrino mass via the seesaw mechanism \cite{Mohapatra:1979ia}. 

Currently, the most sensitive experimental method to establish that neutrinos are Majorana particles is the search for neutrinoless double beta decay (\bbonu). This is a hypothetical, very rare nuclear transition in which a nucleus with Z protons decays into a nucleus with Z+2 protons and the same mass number, A, emitting two electrons that carry essentially all the energy released (\Qbb). While the two-neutrino mode of the double beta decay has already been measured in a number of isotopes, the zero-neutrino mode remains unobserved.

The experimental signature of a neutrinoless double beta decay are two electrons with total kinetic energy equal to \Qbb. For this reason, to be able to make the measurement, an experiment must be optimised simultaneously for energy resolution and the rejection of non-signal events with energies similar to \Qbb. Energy resolution is key not only to limit the number of events caused by natural radiation which enter the region of interest but to limit the pollution from the intrinsic background caused by the two neutrino mode. The signature of a \bbtnu\ event only differs from signal in the sum of the energies of the two electrons and, therefore, improved separation is only made possible by improved energy resolution. Backgrounds from ambient sources can be further reduced through a combination of fiducial cuts and particle identification.

The NEXT experiment seeks to make a first measurement of \bbonu\ in \XE\ using a high pressure gas Time Projection Chamber (TPC) with electroluminescent (EL) read-out. It has been designed to provide good energy and spatial resolution to identify separated tracks and increased ionization (`blobs') at their ends. To this end, it uses two different planes for energy measurement and tracking. Photomultipliers behind the cathode detect the primary scintillation light and allow the energy measurement by detecting the electroluminescent light. An array of silicon photomultipliers (SiPMs) behind the anode provides the spatial information for the topological analysis of the events using the EL light. The \bbonu\ search will be carried out using NEXT-100 which will contain $\sim$100~kg of \XE\ gas at 15~bar and is described in detail in Refs.~\cite{Alvarez:2012haa, Gomez-Cadenas:2013lta}. The first phase of the experiment, called NEW, and deploying 10~kg (and 20\% of the sensors), is currently being commissioned at the Laboratorio Subterr\'aneo de Canfranc. The principle of operation of NEXT-100 and NEW and the necessary know-how has been developed using NEXT-DEMO, a large scale prototype which operated at the Instituto de F\'isica Corpuscular in Valencia (Spain) with $\sim$1.5~kg of natural xenon at a pressure of 10~bar (for a detailed description of the prototype, see Refs.~\cite{Alvarez:2012nd,Alvarez:2013gxa}).

The use of a topological particle identification based on the expected signature of a double electron (signal) event compared to that of a single electron (background) produced by the interaction of high energy gammas is presented here. Using the Monte Carlo simulation of the NEXT-DEMO prototype and data taken with NEXT-DEMO a first demonstration of the power of the method has been made. This involved the comparison of single electron tracks originating from the photoelectric interaction of \NA\ gammas and double electron tracks from the pair production of the 2.614 MeV gamma from \TL. 

The paper is organized as follows. The topological signal and reconstruction algorithms are described in Sec.~\ref{Toposig}. The data analysis is described in Sec.~\ref{Analysis} and the results obtained are presented and discussed in Sec.~\ref{Results}. The paper ends with the conclusion in Sec.~\ref{Conclus}.

%% file: src/TopoSig.tex
\begin{figure}
  \centering
  \includegraphics[width=0.45\textwidth]{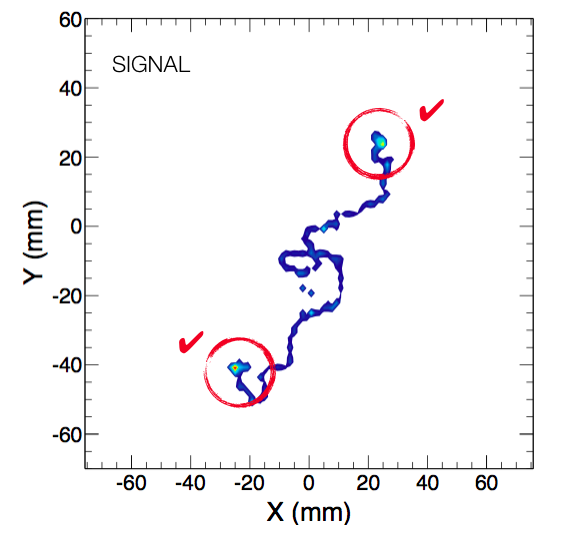}
  \includegraphics[width=0.44\textwidth]{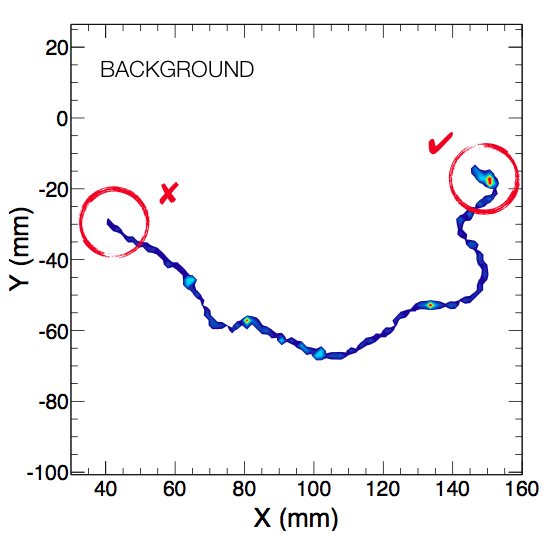}
  \caption{\small A \bbonu\ event (left) and a single electron background event from a 2.44~MeV \BI\ gamma (right) in Monte Carlo simulation (both events simulated at 15~bar gas pressure).} 
  \label{fig.topo}
\end{figure}

Electrons (and positrons) moving through xenon gas lose energy at an approximately fixed rate until they become non-relativistic. At the end of the trajectory the $1/v^2$ rise of the energy loss (where $v$ is the speed of the particle) leads to a significant energy deposition in a compact region, which will be referred to as a `blob'. The two electrons produced in double beta decay events appear as a single continuous trajectory with a blob at each end (Fig.~\ref{fig.topo}-\emph{left}). Background events from single electrons, however, typically leave a single continuous track with only one blob (Fig.~\ref{fig.topo}-\emph{right}).
The use of this topological signature to eliminate background in \bbonu\ experiments was pioneered by the Caltech-Neuch\^atel-PSI Collaboration in the Gotthard Underground Laboratory \cite{Luscher:1998sd}, using a gaseous \XE\ TPC with multiwire read-out, with a fiducial mass of 3.3~kg of \XE\ at a pressure of 5~atm.

\begin{figure}
\centering
\includegraphics[width=0.45\textwidth]{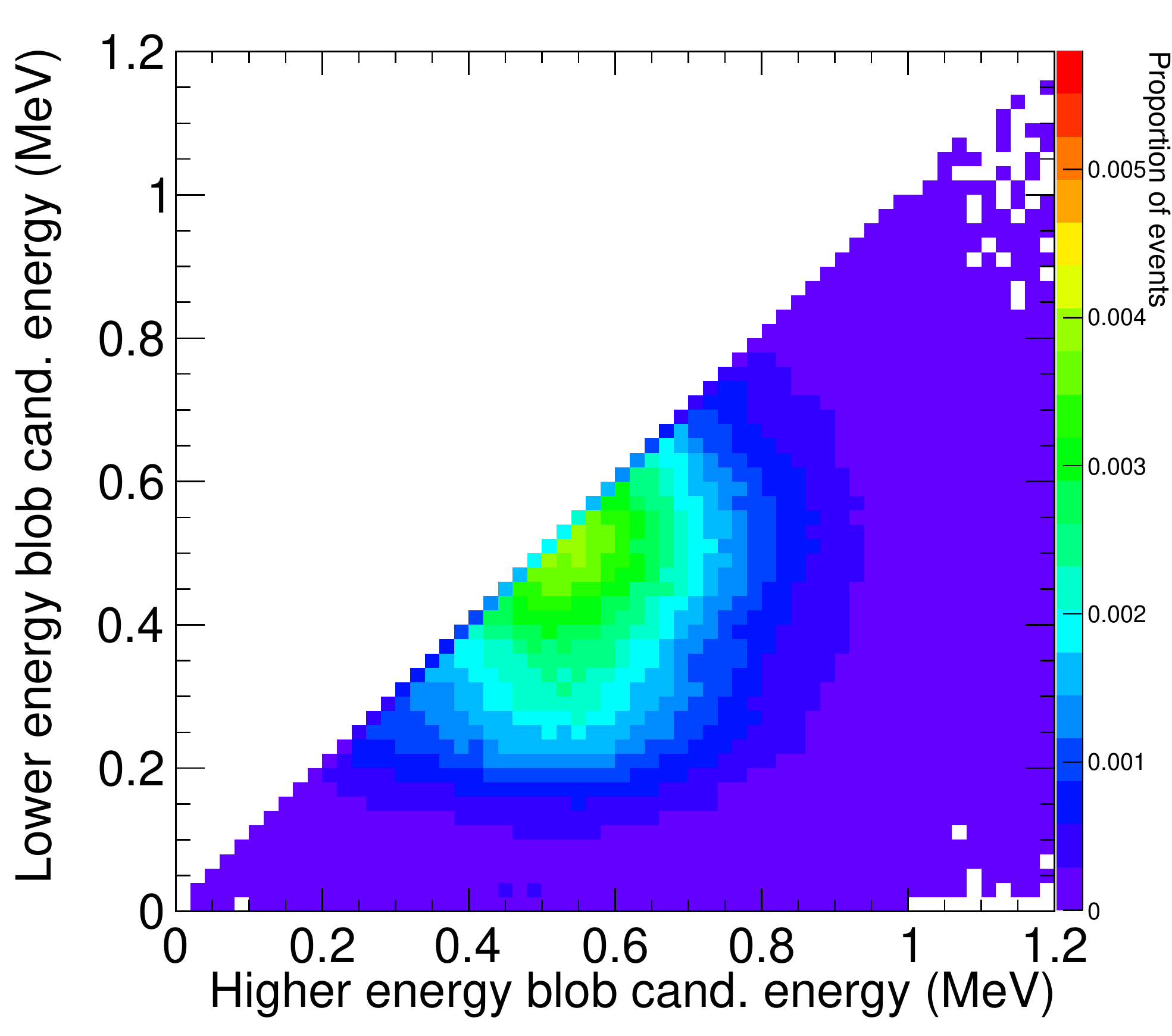}
\includegraphics[width=0.45\textwidth]{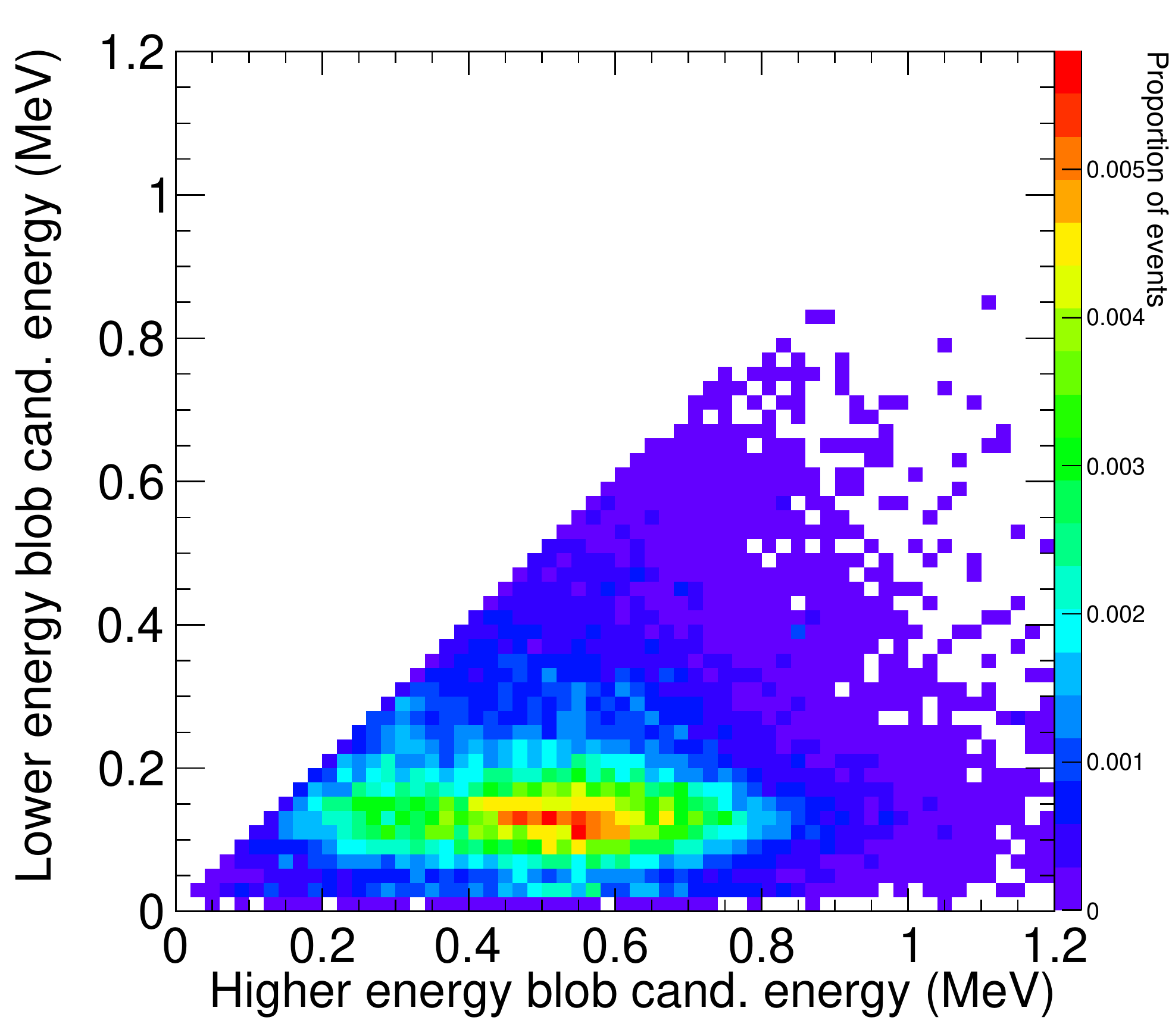}
\caption{\small Distribution of the energy deposition in the two blob candidates of a track, for Monte Carlo \bbonu\ (left) and \TL\ (right) events, simulated with the NEXT final detector \cite{bckgrmodel}. Blob candidates are defined as the sum of all charge within a 2~cm radius of each end-point of the tracks.} 
\label{fig.MC100}
\end{figure}

At the 15~bar pressure of NEXT, the two electrons emitted in a neutrinoless double beta decay tend to leave a single track of about 15~cm length. Energy deposition is approximately constant along the length of the track other than at the two extremes where the electrons deposit more than 20\% of the event energy, split between both blobs. The main background in NEXT comes from high energy gammas emitted in \TL\ and \BI\ decays, which occur naturally in the detector materials as part of the \ensuremath{{}^{232}{\rm Th}} and \UDTO\  chains, entering the active volume of the detector. These gammas convert in the gas through photoelectric, Compton and pair production processes producing, typically, more than one electron separated in space, the tracks of which have just one blob of energy at one extreme. Figure \ref{fig.MC100} shows the distribution of the energy deposited in the blob candidates of the tracks for signal and background events, simulated in NEXT-100. As can be seen, the energies of the blob candidates have closer values for signal events with the blob candidate with less energy at much lower energy for background.

In a previous publication \cite{Alvarez:2013gxa}, only the average position of the track was reconstructed, using the barycentre calculated with the total integrated signal in each SiPM. The work presented here uses a more sophisticated algorithm to reconstruct the topology left in the detector by a charged particle. 


\subsection{Reconstruction of tracks in an EL TPC}
Reconstruction of tracks in an electroluminescent TPC is complicated not only by the diffusion of the charge cloud during drift but also by the nature of the read-out. Scintillation light is produced over the whole width of the EL gap (5~mm in NEXT-DEMO) spreading the signal from a single electron over a time inversely proportional to the drift velocity within the gap ($\sim2$~$\mu$s). Additionally, the EL light is produced isotropically and, therefore, the signal produced by the passage of an electron through the gap is expected to arrive at the tracking plane ($\sim$7.5~mm behind the anode) over the area defined by the intersection of the plane with the sphere of light.

In a previous paper~\cite{Lorca:2014sra}, the NEXT collaboration demonstrated that a `point-like' deposition of charge due to the absorption of a $K_\alpha$ X-ray is expected to be detected over a transverse region which can be parameterised as a two dimensional Gaussian with a standard deviation of $\sim$8~mm where the spread due to EL light production is the dominant effect with subdominant contributions from transverse diffusion of the charge. Longitudinally, the expected spread has a noticeable dependence on the drift distance since the diffusion dominates. $K_\alpha$ events are expected to have widths in $z$ with standard deviations of between 0.5~mm, for very short drifts, and 1.7~mm at the drift field settings used here. In order to optimise the reconstruction of tracks these values must be taken into account by dividing the signal information into appropriate time slices and using charge information from clustered SiPM channels.

\subsection{Hit-finder algorithm}
\label{Trackreco}
Once a signal-like pulse is selected using the energy plane, the corresponding information for tracking must be analyzed in the SiPM plane. The SiPM noise is mainly due to electronic noise in the laboratory, which is higher than the dark current. The distribution of noise per time sample has been measured for each SiPM through dedicated runs without external signal and used to eliminate from the analysis those time slices which exhibit charge below a certain noise level.

\begin{figure}
\centering
\includegraphics [width=0.49\textwidth]{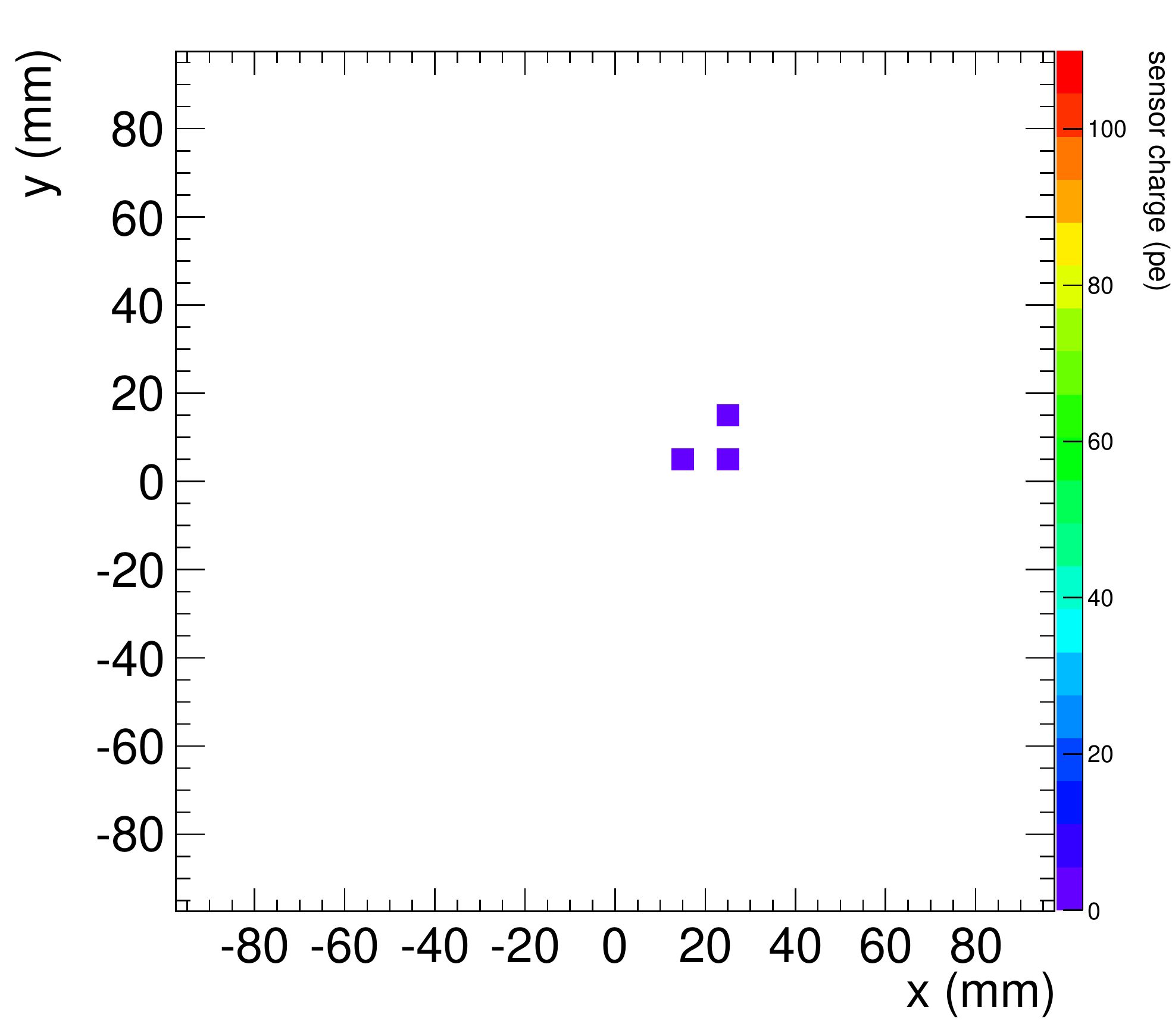}
\includegraphics [width=0.49\textwidth]{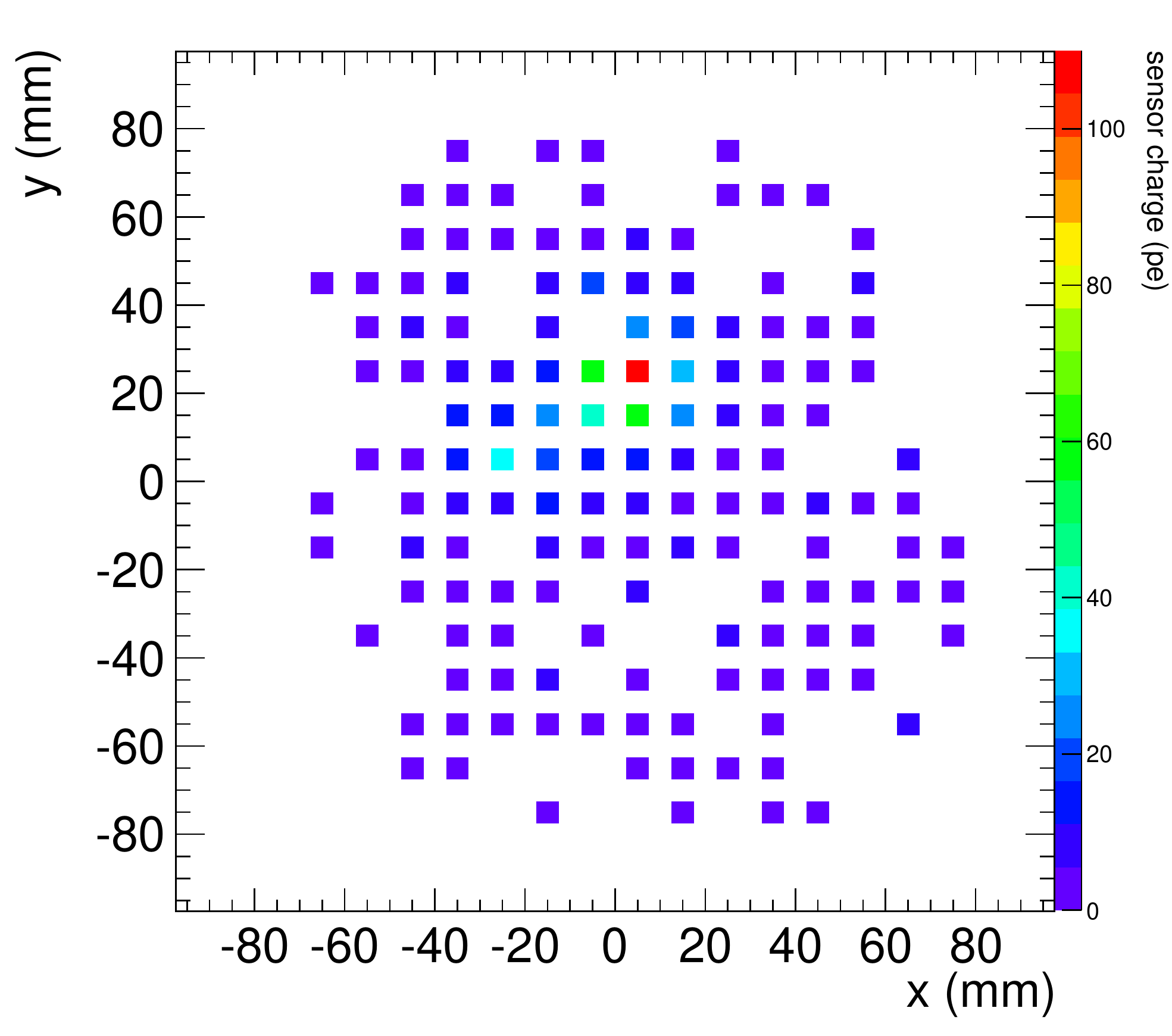}
\caption{\small Tracking plane signals for two arbitrary pulse slices, as an illustration of the hit-finder algorithm. (Left) A slice that does not pass the minimum energy cut, (right) a slice with clear signal (see text for more details).} 
\label{fig.slices}
\end{figure}

The total charge in a time integrated section (slice) of the tracking plane is required to be above a minimum threshold in order to be considered by the algorithm. This requirement has the purpose of eliminating the residual noise, due to fluctuations, without affecting the signal of low energy depositions in the detector. Typically, slices with total charge below threshold are located at the beginning and end of the pulse. In these slices, the position of the sensors with charge is unrelated to the energy depositions reconstructed in the other slices of the pulse. In Fig.~\ref{fig.slices}-\emph{left}, an example of this kind of slice is shown. After this first requirement, the information for each time slice is passed to the hit-finder algorithm.

The algorithm used in this work searches for clusters around local maxima and then proceeds iteratively, selecting first the channel with maximum charge and forming a cluster with the first ring of sensors around it. If, and only if, the ring is fully occupied (taking into account any dead channels or edge effects) and each individual SiPM exhibits a charge higher than a threshold, the cluster information is used to build a hit, whose $x$ and $y$ position are reconstructed as the barycentre of the charge information. The information already used is then removed and the procedure repeated until no SiPM above threshold are left. The need for this threshold can be understood by considering a typical slice located in the centre of a pulse, like the one shown in Fig.~\ref{fig.slices}-\emph{right}. The light seen at the level of a few photoelectrons across the whole SiPM array has undergone multiple reflections before arriving at the sensors and, therefore, provides no relevant position information.
The energy recorded for the corresponding time slice in the energy plane is then divided between all hits found according to the proportion of charge in each hit. This charge is then corrected according to the detector calibration described in Ref.~\cite{Lorca:2014sra}.

\subsection{Voxelization and track identification}
\label{Sec:Voxel}
Once a set of hits is found, a connectivity criterium must be defined so that the hits belonging to each separate particle can be grouped into tracks. The procedure is as follows: first, the active volume is divided into 3D pixels, known as `voxels', of fixed dimensions. Each voxel is given an energy equal to the sum of the energies of all the hits which fall within its boundaries. The collection of voxels obtained in such a way can be regarded as a graph, defined as a set of nodes and links that connect pairs of nodes. Two voxels can then be considered connected if they share a face, an edge or a corner, with each pair of connected voxels being given a weight equal to the geometric distance between their centres. Next, the ``Breadth First Search'' (BFS) algorithm (see, for instance, Ref. \cite{Cormen}) is used to group the voxels into tracks and to find their end-points and length. The BFS algorithm is a graph search algorithm which finds the minimum path between two connected nodes, starting from one node and exploring all its neighbours first, then the second level neighbours and so on, until it reaches the second node. The BFS algorithm divides the voxels into connected sets, known as tracks and finds their end-points, defined as the pair of voxels with largest distance between them, where the distance of two voxels is the shortest path that connects them. The distance between the end-points is the length of the track.


The choice of the voxel size is a compromise between a fine granularity and conservation of connectivity, which depends on the hit-finder algorithm in use. In this work, the best performance has been found for voxels of $1\times1\times1$ cm$^3$. Improvements in the hit-finder algorithm will allow for smaller voxels or different connectivity criteria and a more accurate reconstruction of tracks will be possible. However, one has to take into account that the current transverse and longitudinal diffusion in NEXT-DEMO, as well as transverse spread due to the EL production, affect the position of the ionization electrons during the drift, blurring the track, thus spatial resolution better than the typical diffusion does not provide sufficient additional information for the additional complexity and cost to be worthwhile.  


%% file: src/Analysis.tex
The analysis was performed using data from two different calibration sources to study the topology of background-like and signal-like events in NEXT-DEMO. The active volume of the detector has a drift length of 30 cm and a hexagonal cross section of 8-cm apothem. Background-like events were studied using a 1-$\mu$Ci \NA\ calibration source. The de-excitation of the first excited state of the \NA\ daughter isotope, \ensuremath{^{22}}Ne, produces a gamma with an energy of 1.275~MeV. These gammas can produce a photo-electron which leaves a track in NEXT-DEMO of approximately 7~cm length (10~bar pressure). This track is contained in the TPC and is long enough to perform topological studies.

Signal-like events were obtained using a \THO\ source. The \THO\ decay chain includes \ensuremath{{}^{208}\rm{Tl}}, the daughter of which, $^{208}$Pb, is created in an excited state which de-excites emitting a 2.614~MeV gamma. This gamma can produce an electron-positron pair with a signature which mimics, except for the total energy, the topology of a \bbonu\ event. The positron produces a blob equivalent in energy to that of the electron, and, when it annihilates, emits two back-to-back 511~keV gammas. Due to the size and the pressure of NEXT-DEMO, there is a high probability for both gammas to escape the active volume. In this case, the energy deposited in the chamber is 1.592~MeV and the track left by the electron-positron pair is around 6~cm long. Although the energy of these signal-like events is higher than that of background-like events, the track is slightly shorter on average because both the electron and the positron, individually, have a lower energy than the photo-electron of the \NA\ decay.  These two sources produce tracks of comparable size and energy hence providing two data sets with good characteristics for a topological study.

%% file: src/Data.tex
The \NA\ and \TL\ data were taken under the following detector conditions: 
10~bar pressure, 667~V~cm$^{-1}$ drift field and 2.4
kV~cm$^{-1}$~bar$^{-1}$ EL field. The sources were placed
outside the lateral port positioned midway between cathode and
anode.

The \NA\ data were taken in November 2013 using a trigger on the
scintillation signal optimised for the higher energy gamma compared to
that described in \cite{Alvarez:2012nd}. An integrated exposure of $\sim$84.4~hours
was used in the analysis presented. The \TL\ data were collected in May 2014 using a trigger on the EL
 signal. An event was required to contain at least one ionization
 signal which registered at least 21\,000~photo-electrons (pe) in the central PMT. An
 integrated exposure of $\sim$165.68~hours was used in this case.

Dedicated Monte Carlo (MC) simulation samples were generated using NEXUS, a
Geant4 \cite{Agostinelli:2002hh,Allison:2006ve} based package
developed by the NEXT collaboration that simulates the entire signal generation process. Samples of interacting gammas at the
energy of that from \NA\ and that from \TL\ have been used for event
selection studies. In order to study a possible source of background coming from cosmic muons, an additional sample of muon events with an energy
of 4 GeV and an angular distribution following $\sim$ $\cos^2(\theta)$,
according to Ref.~\cite{Agashe:2014kda}, where $\theta$ is the zenith angle, has been analyzed.

%% file: src/Selection.tex
Charge pulses are identified as originating from primary scintillation (S1) or EL (S2) based on the criteria that they have a duration in time of between 1 and 3~$\mu$s for S1 (\textgreater 3~$\mu$s for S2), and that they have integrated charge of $\geqslant$0.5~pe ($\geqslant$10~pe for S2) in the average of all the PMTs. In this way both the higher energy events of interest for the analysis and the xenon X-rays are included. X-ray pulses allow for the energy scale to be easily validated between separate runs. Only events with one, and only one, S1-like signal are accepted as potential signal to avoid event pile up and for ease of calculation of the $z$ coordinate. Before the data are passed to the hit-finder algorithm, the pulses identified in the events must first pass a number of pre-selection requirements designed to remove any spurious charge due to instrumental effects. These include requirements on the drift to ensure that no S2 has deposited charge within 2~cm of either the cathode or anode; on the relative rms between the recorded signal in the PMTs rejecting S2s with PMT rms greater than 20\% of the mean signal; on the ratio between anode and cathode charge; and a loose transverse fiducial requirement on the integrated SiPM charge ($|x|$ and $|y| \leqslant 60$~mm). Table~\ref{tab:effSum} shows the effect of this pre-selection on the data and Monte Carlo samples. The table also shows the reductions in the total data samples for the two sets. The pre-selection requirements remove far more data events since they are designed to remove events that are not induced by the source which are not present in the MC sample, such as interactions in the cathode buffer region and small sparks. However, it can be seen that at the level of the final requirement the samples are at a comparable level for the \NA\ data with the difference seen for the \THO\ sample explained by the presence of single electron events in the region of interest (ROI) as described in Sec.~\ref{subSec:DEMORes}, validating the filtering method.

All events which passed the preselection were then presented to the hit-finder algorithm described in Sec.~\ref{Trackreco} and subjected to a final fiducial requirement based on the position of the reconstructed hits. No event may have any hits reconstructed outside the cylinder with cross section defined by $R = \sqrt{x^2 + y^2} = 50$~mm and $z$ limits as defined by the fiducial cut described above ($20<z<280$ mm). This requirement ensures that entering backgrounds and events only partially contained are removed from the data set. All remaining simulated cosmic muon events were rejected by this fiducial requirement.

\begin{table*}[h!]\centering

\begin{tabular}{@{}l|cc|cc@{}}\toprule
  & \multicolumn{4}{c}{Remaining sample proportion}\\
  Selection & \multicolumn{2}{c}{MC} & \multicolumn{2}{c}{Data}\\
             & \TL\ & \NA\  &  \TL\ & \NA\ \\  \midrule
  Pre-selection & 0.451$\pm$0.002  & 0.0769$\pm$0.0001 & 0.329$\pm$0.001 & 0.0608$\pm$0.0002 \\
  xy inside fid. veto &  0.244$\pm$0.002 &  0.411$\pm$0.001 & 0.286$\pm$0.001 & 0.316$\pm$0.002 \\
  Energy in the ROI & 0.470$\pm$0.005 & 0.0071$\pm$0.0002 & 0.274$\pm$0.003 & 0.063$\pm$0.002\\
  One track identified & 0.955$\pm$0.003& 0.976$\pm$0.005 & 0.887$\pm$ 0.004 & 0.847$\pm$0.010 \\
  Two blobs found & 0.658$\pm$0.007 & 0.219$\pm$0.013 & 0.558$\pm$  0.006 & 0.243$\pm$0.014 \\
  
 \bottomrule
\end{tabular}
\caption{Event selection efficiencies for each cut requirement described in the text evaluated for signal MC plus muon sample and for data. Errors are statistical.}\label{tab:effSum}
\end{table*}

\begin{figure}[htb]
  \begin{center}
    \includegraphics[width=0.49\textwidth]{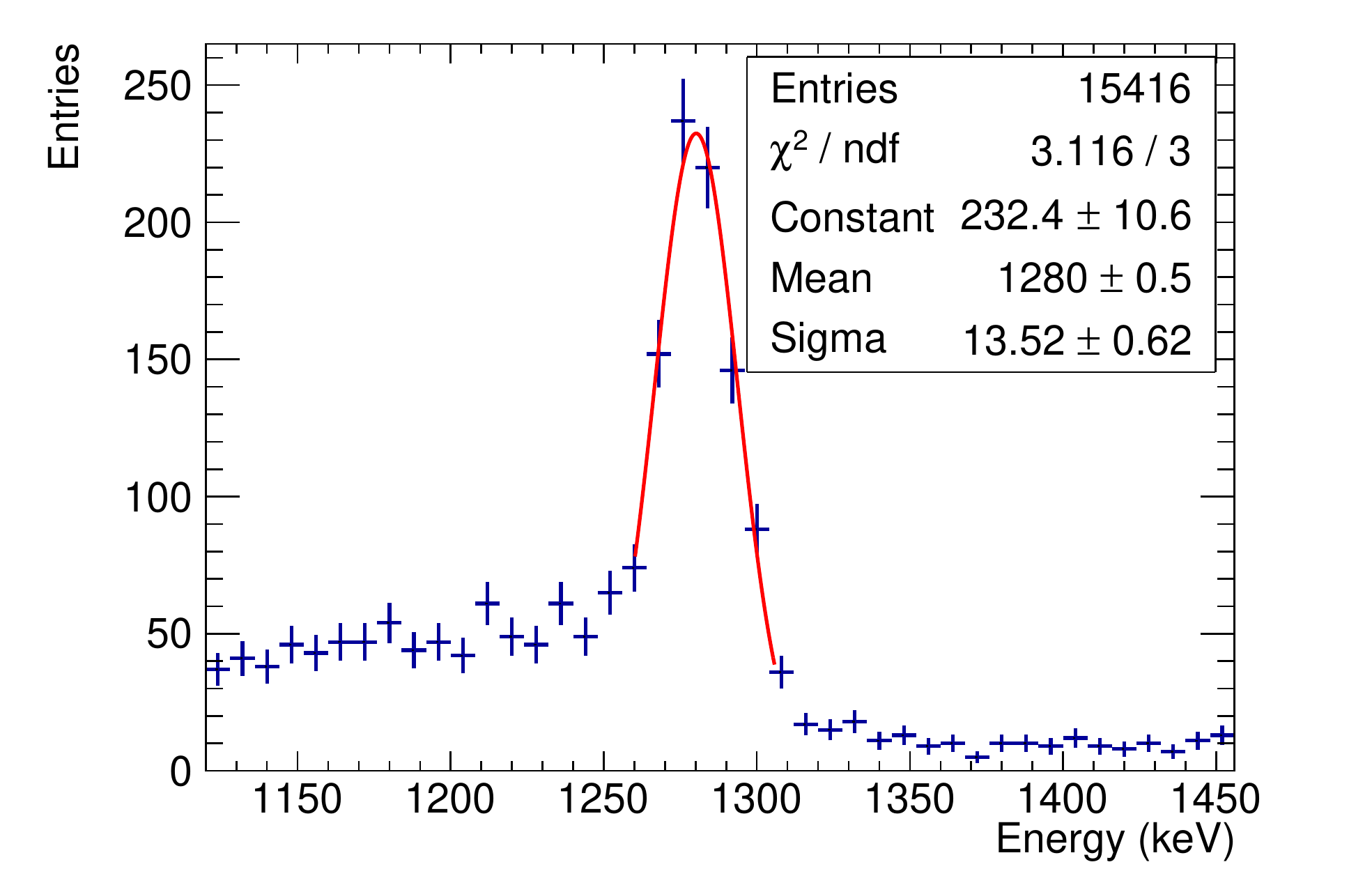} 
    \includegraphics[width=0.49\textwidth]{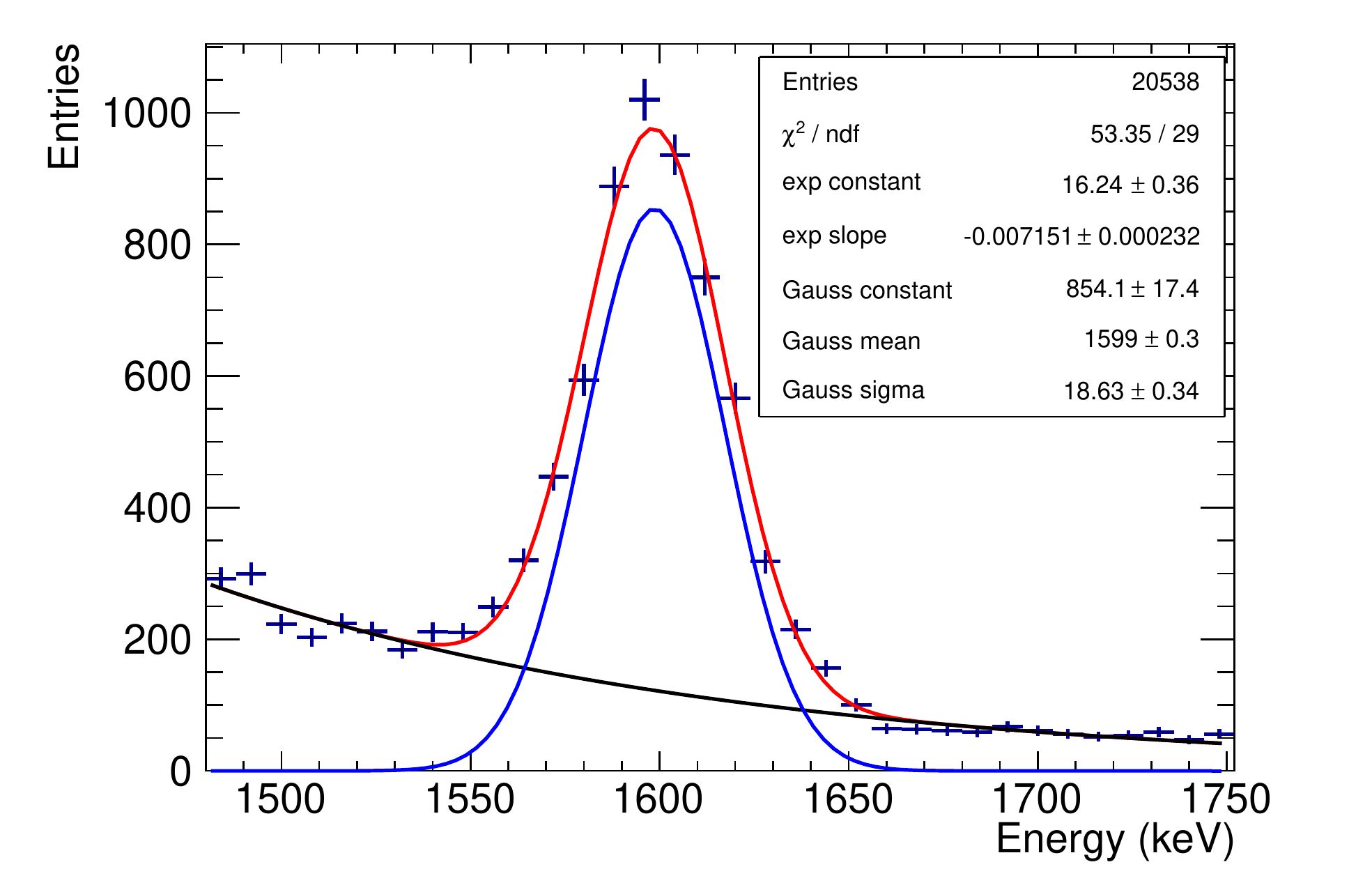}
    \caption{\small Energy spectra for the \NA\ source (left) and the \THO\ source (right), for one-track only events passing the selection criteria described in the text.}
    \label{fig:fits}
  \end{center}
\end{figure}

After the final fiducial requirement the energy of each S2 was reconstructed using the position dependent calibration described in Ref.~\cite{Lorca:2014sra}. Events were then voxelised and tracks were constructed using the algorithms described in Sec. \ref{Sec:Voxel}. Using voxels of $1\times1\times1$~cm$^3$, track sections were considered in three dimensions and, when two reconstructed tracks of at least four voxels were separated by at most one voxel, they were merged to form a single track. Figure \ref{fig:fits} shows the energy spectra for both sources for all events with one merged track (details on the one-track requirement will be discussed later in the text). In these figures, the photoelectric peak of the interaction of the \NA\ 1.275~MeV gamma and the double escape peak resulting from pair production induced by the 2.614 MeV gamma can be seen. The \TL\ data were taken 3 months after the calibration run resulting in a poorer energy resolution. In NEW and NEXT-100 regular calibration runs will be taken to avoid similar decalibration.

In order to exclude events which are not photoelectric interactions (in \NA\ data) or pair production events  (in the case of \TL) a ROI was selected for each dataset. The \NA\ ROI was chosen by fitting a Gaussian to the 1275~keV peak and considering all events with energy within $\pm$3~$\sigma$ of the mean obtained from the fit. A similar selection was made for \TL\ data taking into account that in the region of the double escape peak single electron events from Compton scattering are also expected. The fit was performed parameterising the fall of the Compton edge with an exponential and the signal with a Gaussian distribution. The fit predicts that 25.7 $\pm$ 0.6\% of events are single electrons and 74.3 $\pm$ 0.6\% are pair production events in a region within $\pm$3~$\sigma$ of the mean obtained from the fit. The exponential background model fits well in the regions immediately before and after the peak, and the variation of the range used in the combined fit predicts errors on the signal to background ratio of less than 1\%.

Those events which have tracks with energies within the ROI are then filtered according to the number of tracks identified. Events with more than one track are not selected for the final sample. At the energies considered here both signal and background are expected to produce only one track due to which this requirement has little effect. However, in the \bbonu\ analysis described in Ref.~\cite{bckgrmodel} selecting only one track proves important and is kept here for completeness. As can be seen in Table~\ref{tab:effSum}, 89\% of the \TL\ double escape peak candidates and 85\% of the \NA\ 1275~keV gamma candidates fulfill this requirement.

\begin{figure}[htb]
\begin{center}
\includegraphics[width=0.49\textwidth]{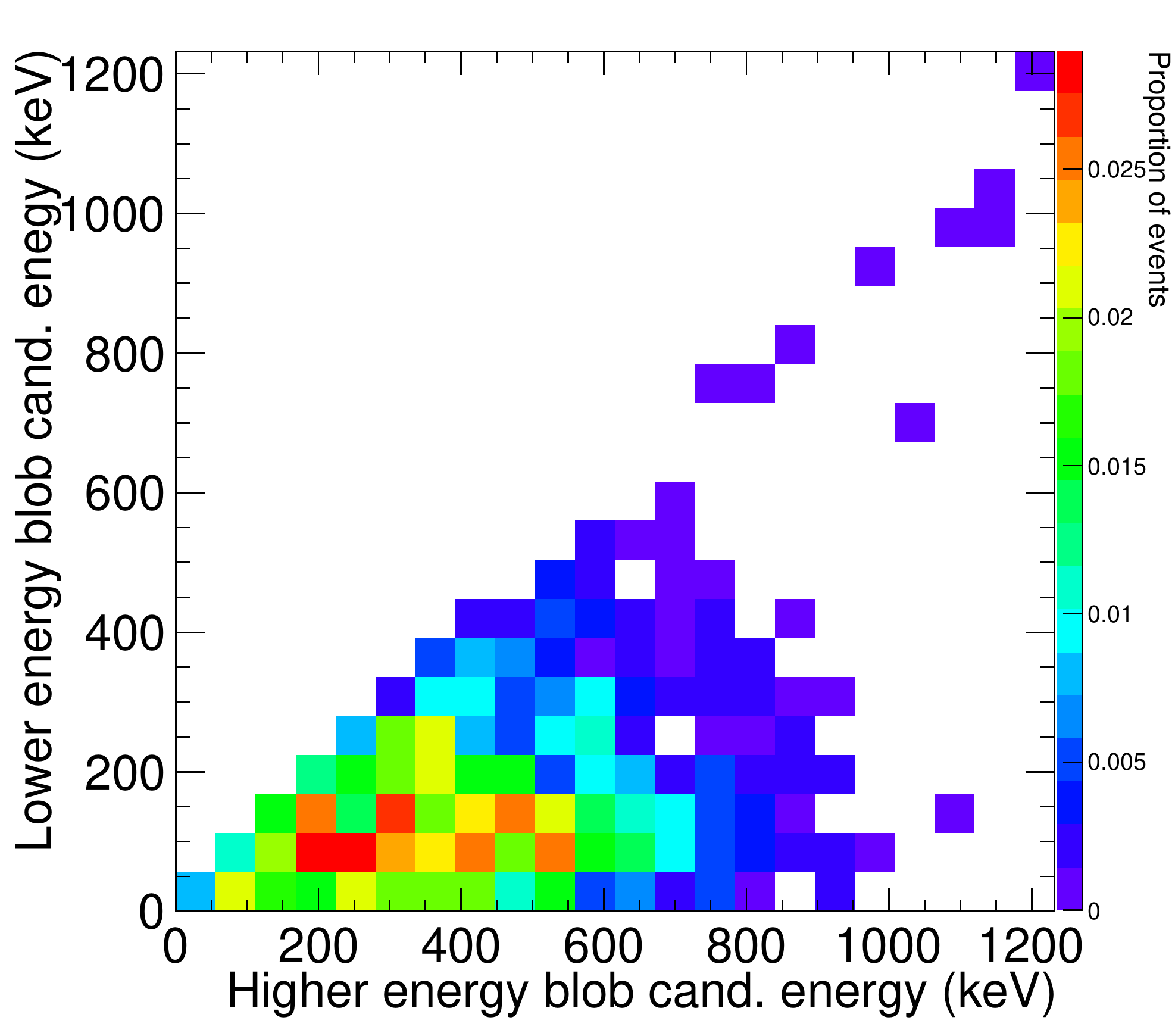} 
\includegraphics[width=0.49\textwidth]{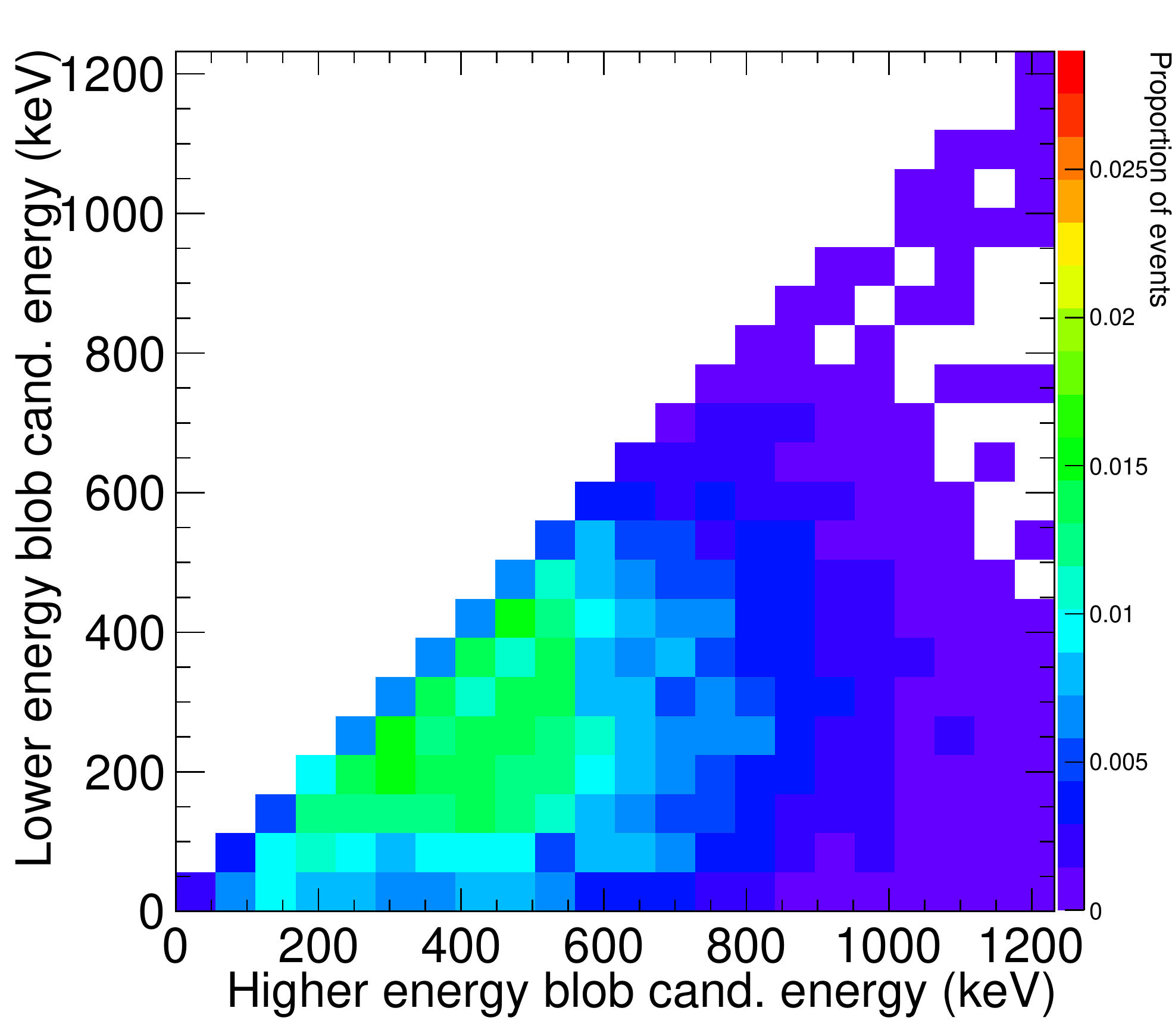}
\caption{\small Energy distribution at the end-points of the tracks coming from \NA\ decay (left) and those coming from the \TL\ decay (right) for 2~cm radius blob candidates.}
\label{fig:2blobs}
\end{center}
\end{figure}

For each track in the ROI, the blobs were sought at their end-points. A blob candidate is defined as
 a group of voxels which fall within a set radius from a track end-point. Figure \ref{fig:2blobs} shows, for both datasets, the energies of the blob candidate with lower energy plotted against that of the other blob candidate. It can be seen that the single-electron-dominated sample from \NA\ tends to have less energy in its lower energy blob candidate. A slight shift in the energies of the higher energy blob candidates is seen also, which is expected to be due to calibration effects.

%% file: src/Results.tex
\begin{figure}[htb]
  \begin{center}
    \includegraphics[width=0.9\textwidth,]{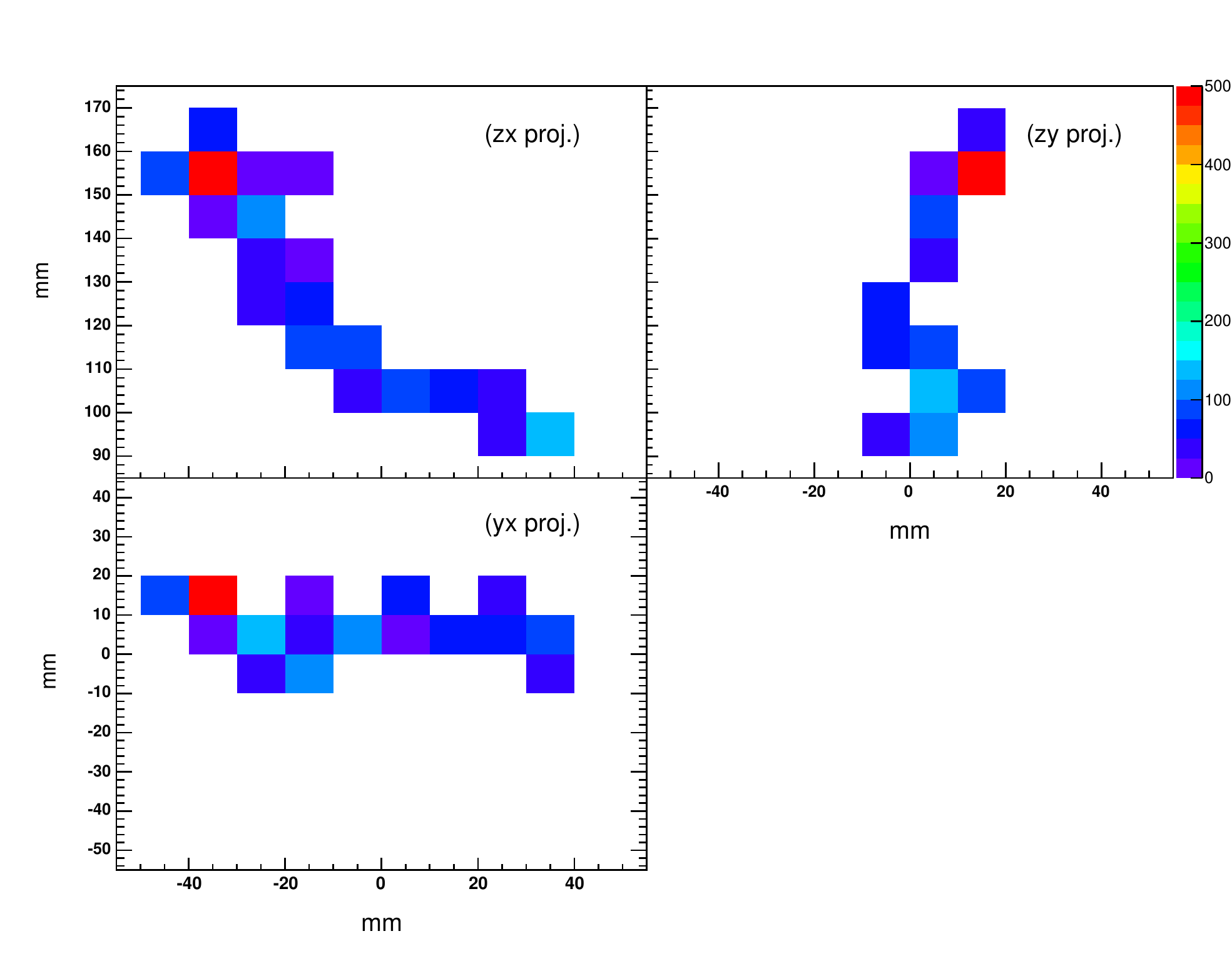} 
    \caption{\small Example of a \NA\ gamma track. The $zx$ and $zy$
      projections (on top) and the $yx$ projection in the lower
      frame are drawn. The energy scale is in keV (color coded) and
      the distances are in mm. The single electron candidate exhibits one blob
      at one end of the track.}
    \label{fig:NaTrack}
  \end{center}
\end{figure}
\begin{figure}[htb]
  \begin{center}
    \includegraphics[width=0.9\textwidth,]{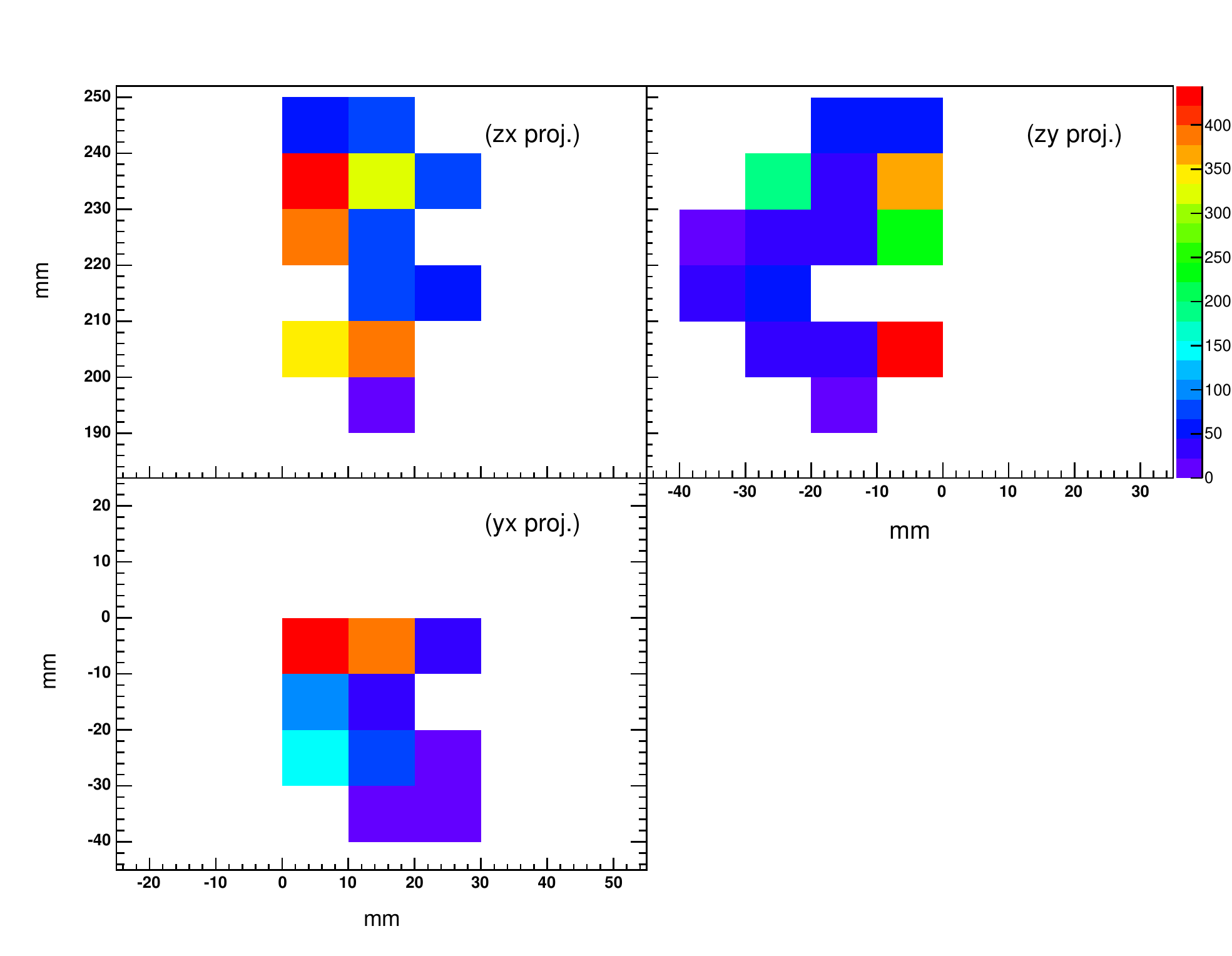} 
    \caption{\small Example of a double escape peak candidate from the
      \TL\ data sample. The $zx$ and $zy$
      projections (on top) and the $yx$ projection in the lower
      frame are drawn. The energy scale is in keV (color coded) and
      the distances are in mm. The double electron candidate exhibits two blobs, one at each end of the track.}
    \label{fig:ThTrack}
  \end{center}
\end{figure}

Single electrons are only expected to have a high energy deposition at
one end of their track as opposed to double electrons which are
expected to have large energy
depositions at each end of the track. An example of single electron track
candidate is shown in Fig.~\ref{fig:NaTrack} and a typical example
of double electron track candidate is shown in Fig.~\ref{fig:ThTrack}.
Figure \ref{fig:2blobs} shows that a requirement on the energy of the less energetic blob candidate is the
simplest criterium to separate signal (\TL\ e$^+$e$^-$) from
background (\NA\ e$^-$) events. Optimisation of this requirement uses the standard
figure-of-merit $\epsilon/\sqrt{b}$, where $\epsilon$ is the signal
efficiency and $b$ is the fraction of background events that survives the requirement in the signal region.

\subsection{Monte Carlo studies}
\label{subSec:MCVer}
\begin{figure}[htb]
  \begin{center}
  \includegraphics[width=0.49\textwidth]{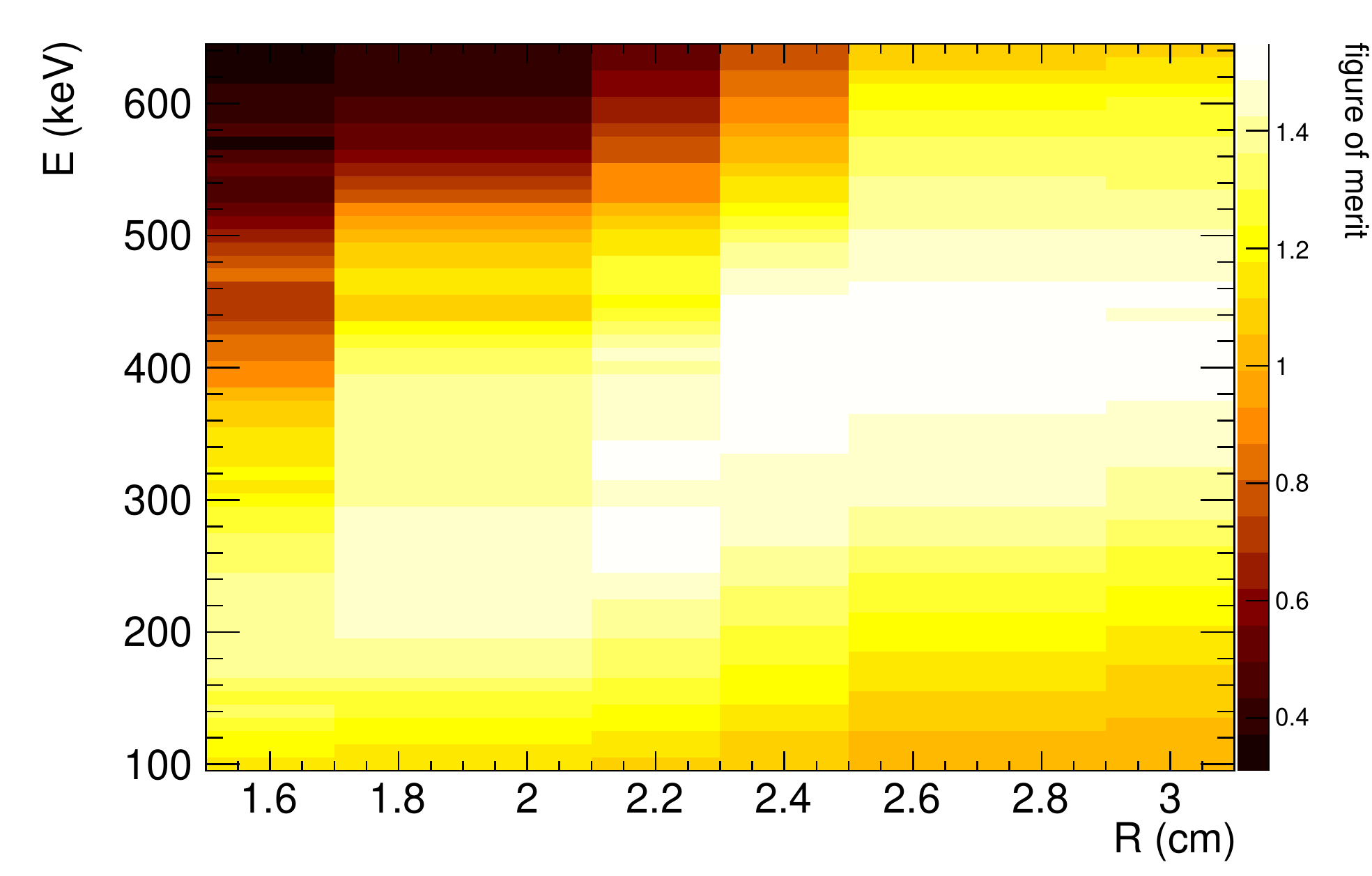} %
  \includegraphics[width=0.49\textwidth]{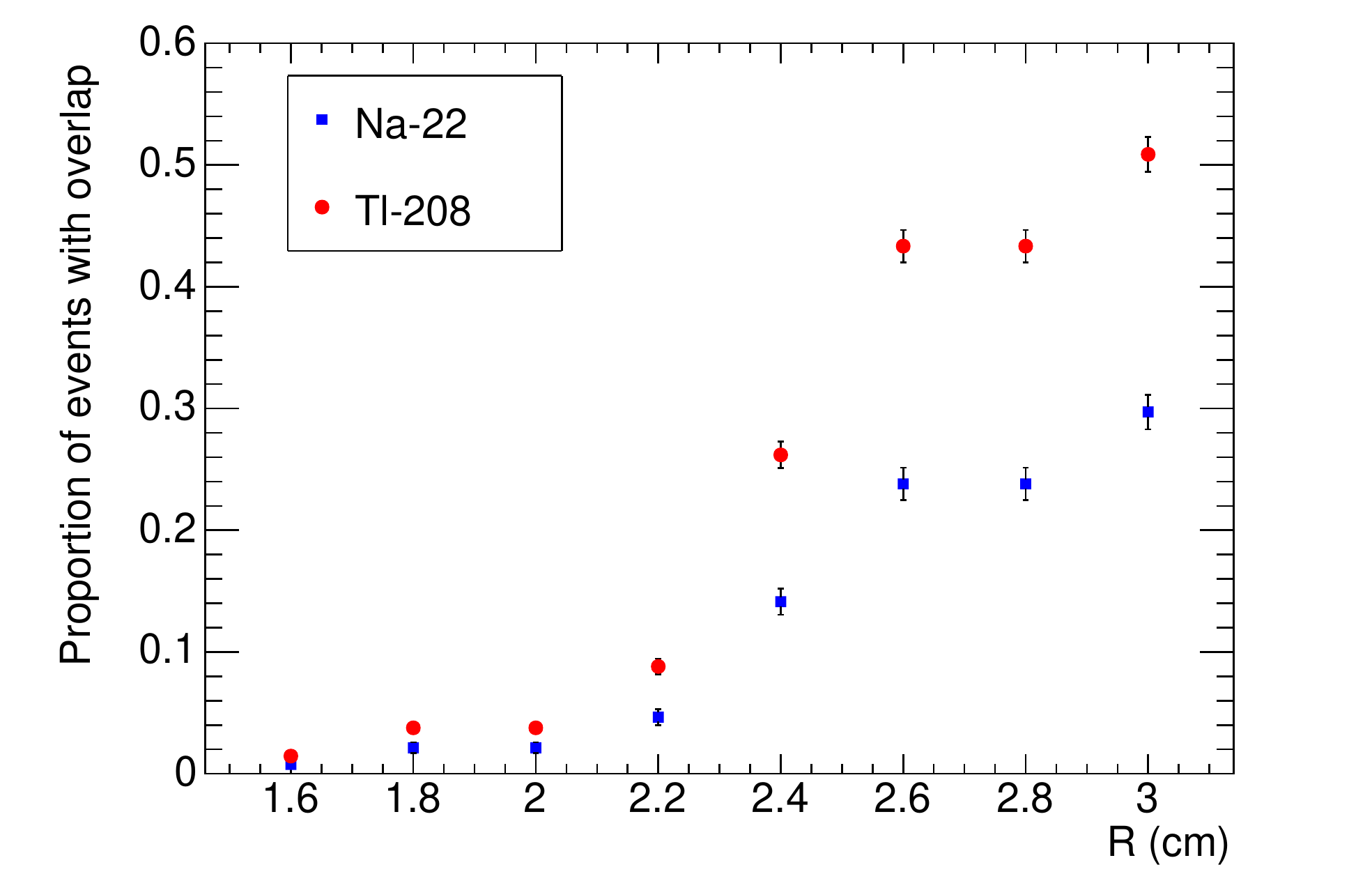} %
    \caption{\small Left: the figure of merit, $\epsilon/\sqrt{b}$ as a function of the blob candidate radius and the minimum blob candidate energy. Right: proportion of events with blob candidate overlapping, as a function of the blob candidate radius. The step-like behaviour that appears in both plots is an artificial consequence of the 1-cm size of voxels. For certain values of radius, the number of voxels inside the blob candidate, thus the proportion of events with overlap, does not change.}
    \label{fig:fom}
  \end{center}
\end{figure}

The evaluation of the figure of merit was performed as a function of
the blob candidate radius and the minimum energy required for the low energy
blob candidate, using MC data (see Fig.~\ref{fig:fom}-\emph{left}). An
additional consideration is the probability of blob candidate region overlap. In
Fig.~\ref{fig:fom}-\emph{right}, the fraction of events with
overlapping blob candidates is shown for different radii, for both \NA\ and \TL\
samples. For larger radii, blob candidate overlap occurs in a higher percentage
of events, particularly in the case of signal events which tend to be
shorter. A radius of 2~cm is found to be optimal if one considers that it is large enough
to contain the energy deposition at the end of the tracks while being small enough,
compared to the average track length, to limit the number of tracks in
which voxels are considered in both blob candidates to $\sim$3\%. For blob candidates of
2~cm radius, an optimum requirement on the minimum blob candidate energy is
found at 210~keV, as can be seen in
Fig. \ref{fig:fom}-\emph{left}. From the range of values that give
very similar figures of merit, that with higher signal efficiency was selected.

\begin{figure}[htb]
\begin{center}
\includegraphics[width=0.49\textwidth]{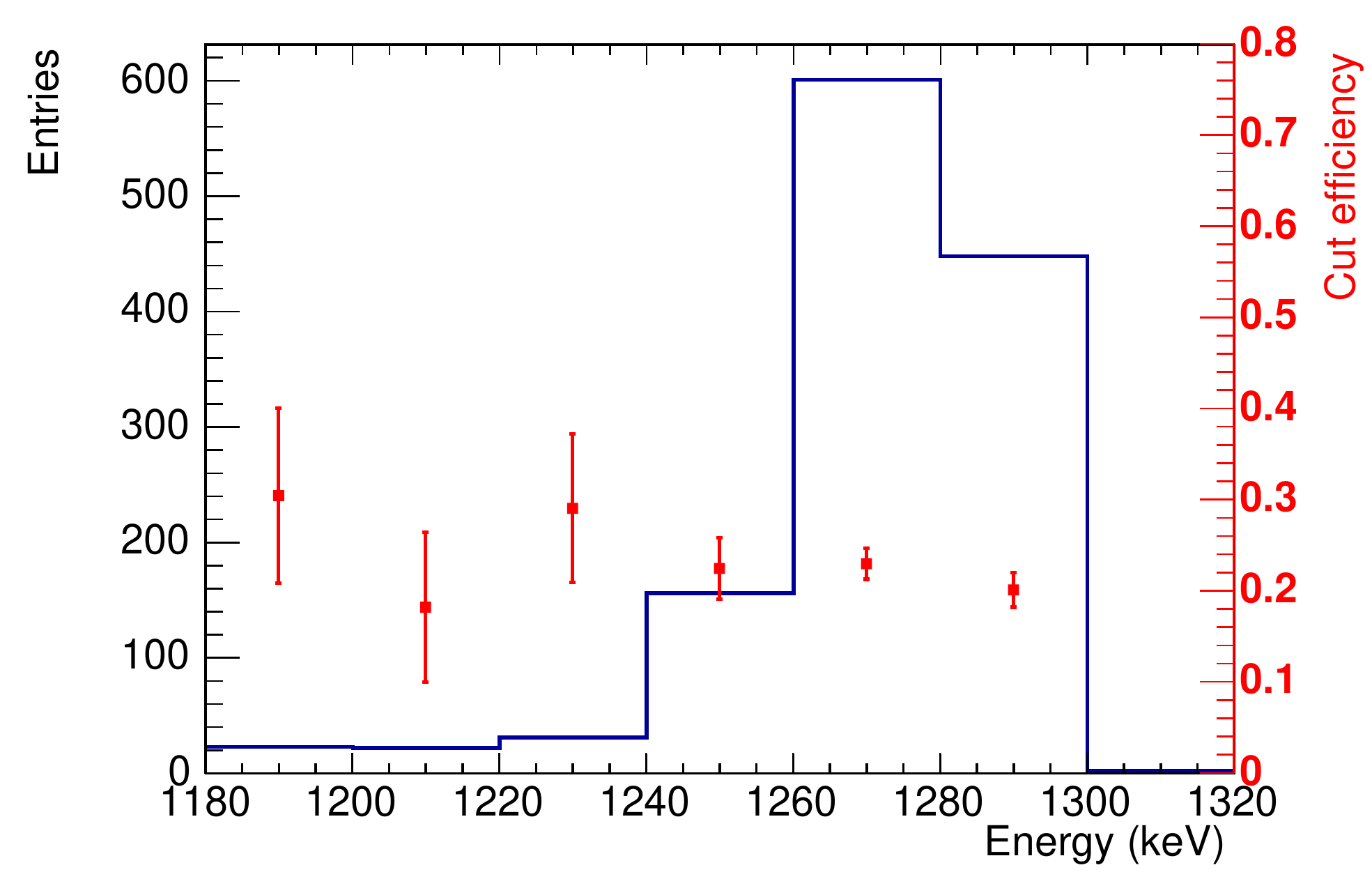} %
\includegraphics[width=0.49\textwidth]{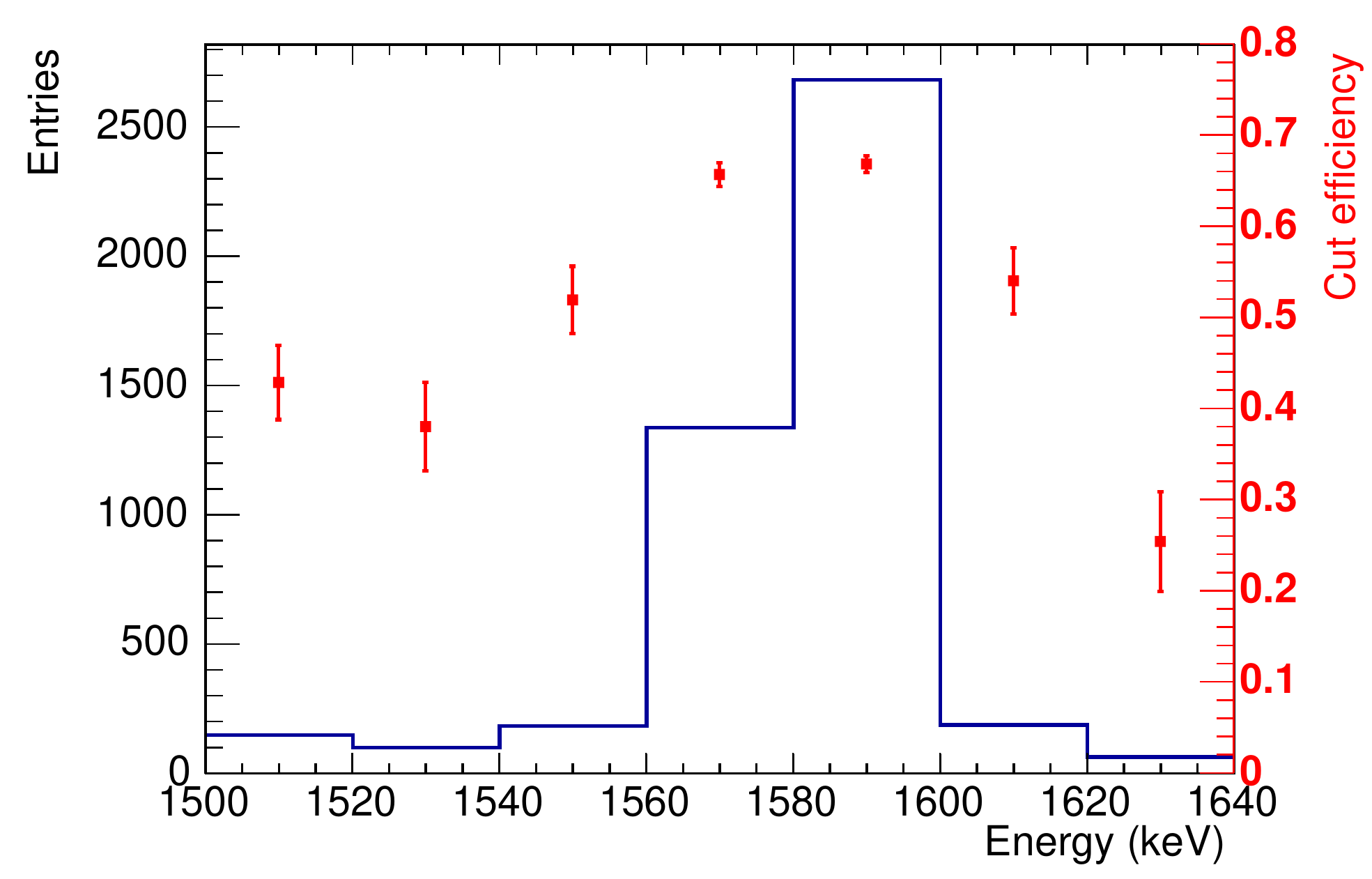} 
\caption{\small Two-blob cut efficiency for the Monte Carlo \NA\ (left) and \TL\ (right) samples. The histograms represent the energy of the event, while the points are the efficiency of the two-blob cut for the events in the bin.}
\label{fig:EffMCall}
\end{center}
\end{figure}

\begin{figure}[htb]
  \begin{center}
    \includegraphics[width=0.49\textwidth]{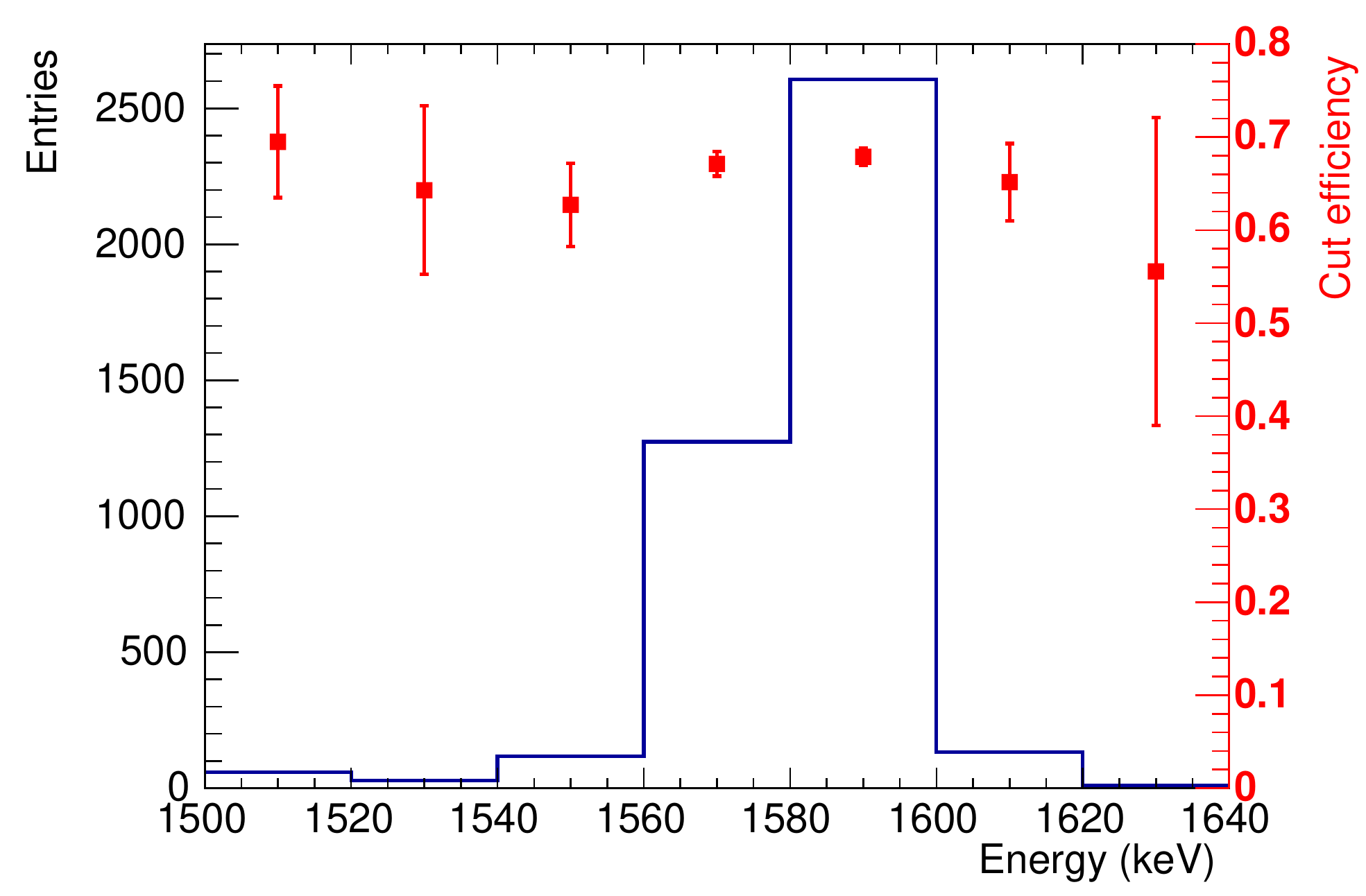} 
    \includegraphics[width=0.49\textwidth]{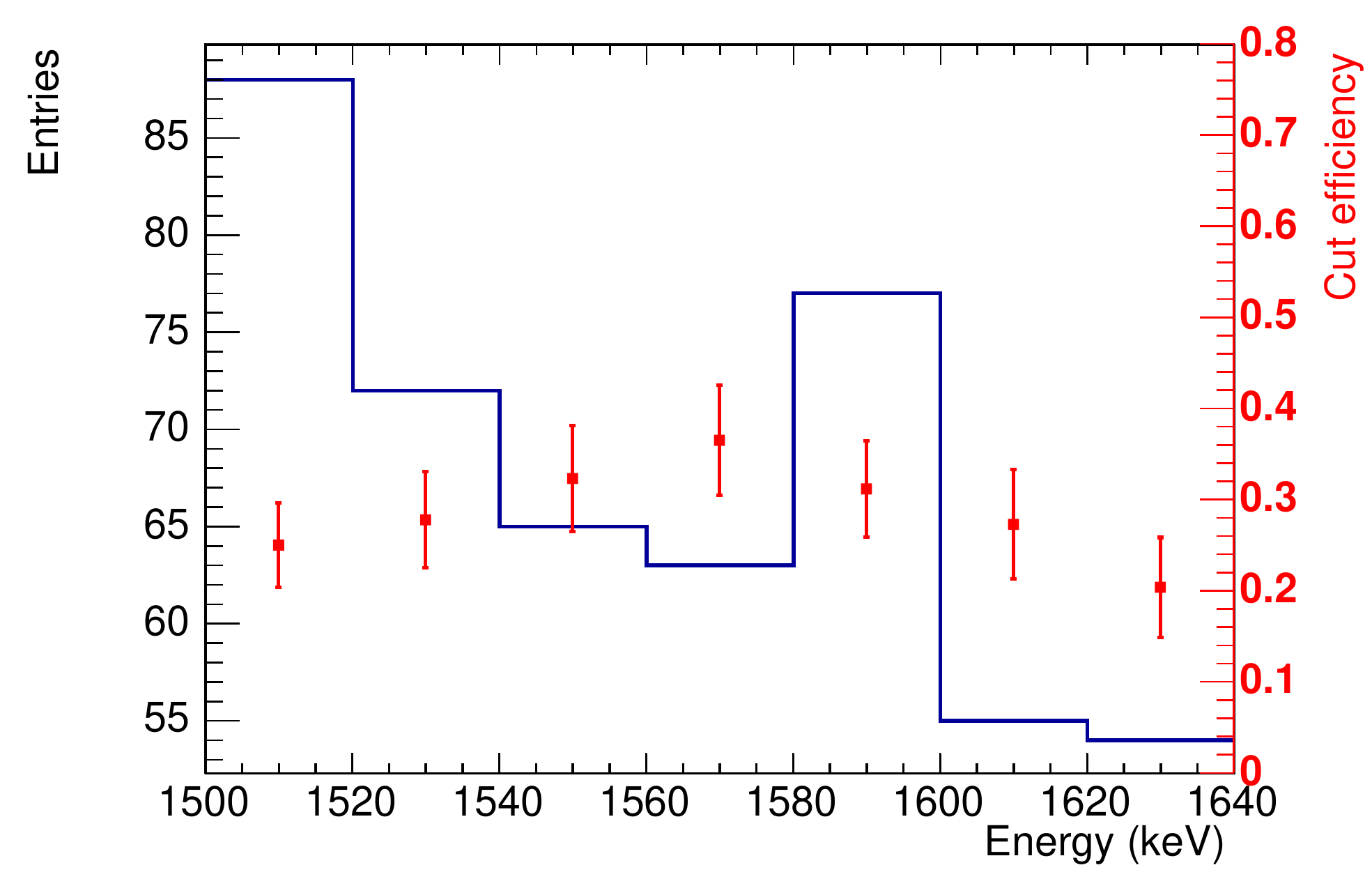} 
        \caption{\small Two-blob cut efficiency for the Monte Carlo \TL\ sample, selecting events with pair production only (left) and events without pair production (right). The histograms represent the energy of the event, while the points are the efficiency of the two-blob cut for the events in the bin.}
    \label{fig:EfiMC}
  \end{center}
\end{figure}

Applying this requirement to the minimum blob candidate energy allows 65.9\% of the
remaining \TL\ events to pass and 21.9\% of the \NA\ events. 
%
%
Figure~\ref{fig:EffMCall} shows the efficiency for acceptance/survival
of the two blob requirement for the \TL\  and \NA\  samples in the ROI as a
function of the observed energy. Events in the \NA\ sample are
expected to consist of single electrons and the survival efficiency
is not expected to be a strong function of the electron energy in the
small range of the ROI, see Fig.~\ref{fig:EffMCall}-\emph{left}. The final
sample of \TL\ events is expected to be a sum of the pair production
and single electrons from Compton scattering with the composition
dependent on the event energy, therefore it is expected that the
acceptance efficiency should exhibit energy dependence, as
illustrated in Fig.~\ref{fig:EffMCall}-\emph{right}. The efficiency of the
two blob requirement on these two event categories was studied using
separated Monte Carlo event samples, as shown in
Fig.~\ref{fig:EfiMC}. As expected, the relative population of these
categories varies strongly with the event energy, but the effect of the
requirement for a given event category does not vary significantly
within the ROI. The survival factor for the simulated single electron
events in the \TL\ sample is found to be consistent with that
measured for the \NA\ sample, between 20\% $-$ 30\%, whereas the
efficiency of acceptance of double track events is at the level of
65\% $-$ 70\%.

\begin{figure}[htb]
  \begin{center}
 \includegraphics[width=0.49\textwidth]{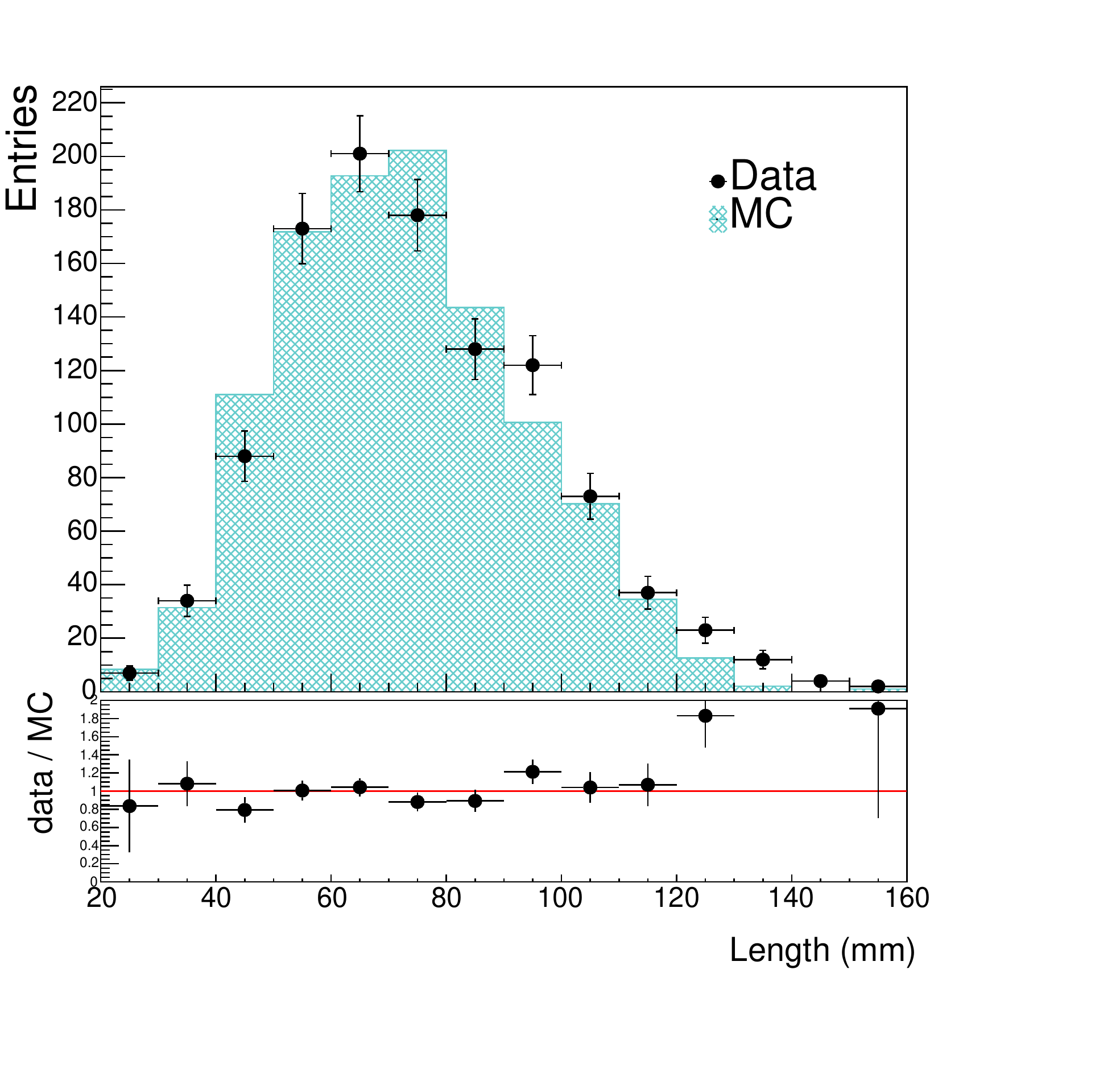} %
    \includegraphics[width=0.49\textwidth]{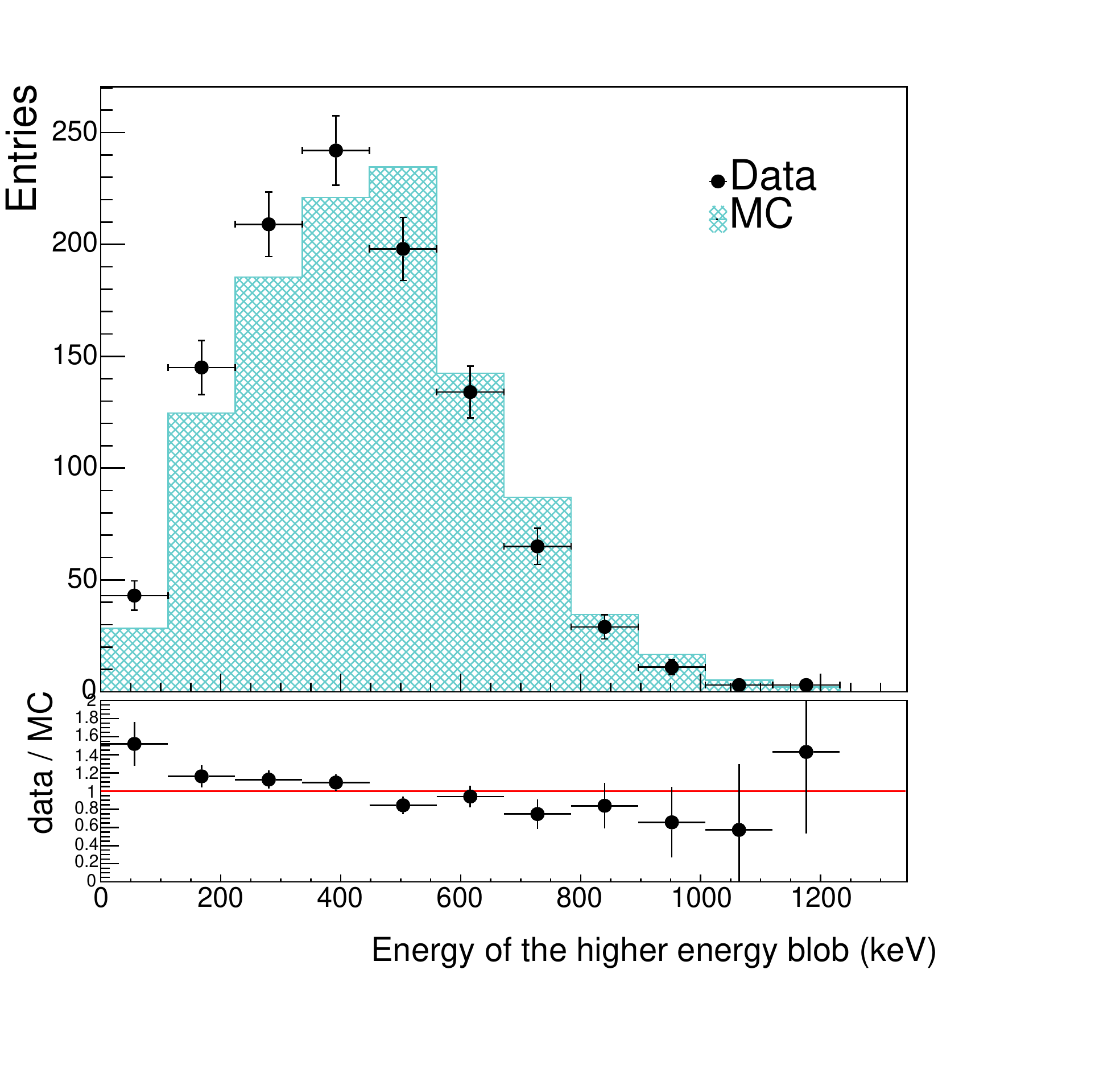} %
    \caption{\small Comparison between data and Monte Carlo
      simulations, for the reconstructed length of the tracks (left)
      and the reconstructed energy inside the blob candidate with
      higher energy (right). \NA\ samples are used. Bins 12 and 13 in
      the bottom left plot are not shown due to lack of MC statistics (larger samples will be used for the future analysis).}
    \label{fig:Lengths}
  \end{center}
\end{figure}

Validation of the MC and reconstruction methods has been performed. 
The MC was found to reproduce the topological features 
found in data to a high degree of accuracy as can be seen in
Fig.~\ref{fig:Lengths}-\emph{left}, for the track
lengths and in Fig.~\ref{fig:Lengths}-\emph{right} for the energy of the
higher energy blob candidate shown for data and MC. In Fig.~\ref{fig:Lengths}-\emph{right} we see a slight systematic shift between MC and data, which we will study in greater detail in the future.


%


\subsection{NEXT-DEMO results}
\label{subSec:DEMORes}
Applying the same selection to the data taken with NEXT-DEMO
results in a proportion of $55.8 \pm 0.9$ (stat.)\% of the \THO\
dataset passing the two-blob requirement and $24.3
\pm 1.4$ (stat.)\% for the \NA\ sample. In the case of \TL\ this result does not
represent an efficiency of identification of double escape peak events
since the ROI contains both single electron tracks (Compton events)
and double electron tracks (pair production events). Assuming the proportions of each interaction type predicted by
the parameterisation of the spectrum described in Sec.~\ref{Selection} and the single-electron survival of the \NA\ data, an
estimation of the signal efficiency can be made according to the
following equality:
\begin{equation}
  \epsilon_{\mathrm{signal}} = \frac {\epsilon_{\mathrm{total}} - \epsilon_{\mathrm{bkg}} \times f_{\mathrm{bkg}}}
  {f_{\mathrm{signal}}} \, ,
\end{equation}
where $\epsilon_{\mathrm{signal}} $ and $\epsilon_{\mathrm{bkg}}$ are
the proportions of signal and single-electron events that pass the cut,
$\epsilon_{\mathrm{total}} $ is the total proportion of events in the
final sample, and $f_{\mathrm{bkg}}$ and $f_{\mathrm{signal}}$ are the
fractions of background and signal-like events in the sample subjected
to the cut ($f_{\mathrm{bkg}}+f_{\mathrm{signal}}=1$). This procedure predicts that $66.7\pm 0.9$ (stat.) $\pm$ 0.3 (fit)\%
of the double escape peak events pass the final cut. The same procedure applied to the MC samples gives a result of $68.6\pm 0.8$ (stat.)\% efficiency for the double escape peak events. 



\begin{figure}[htb]
  \begin{center}
       \includegraphics[width=.49\textwidth]{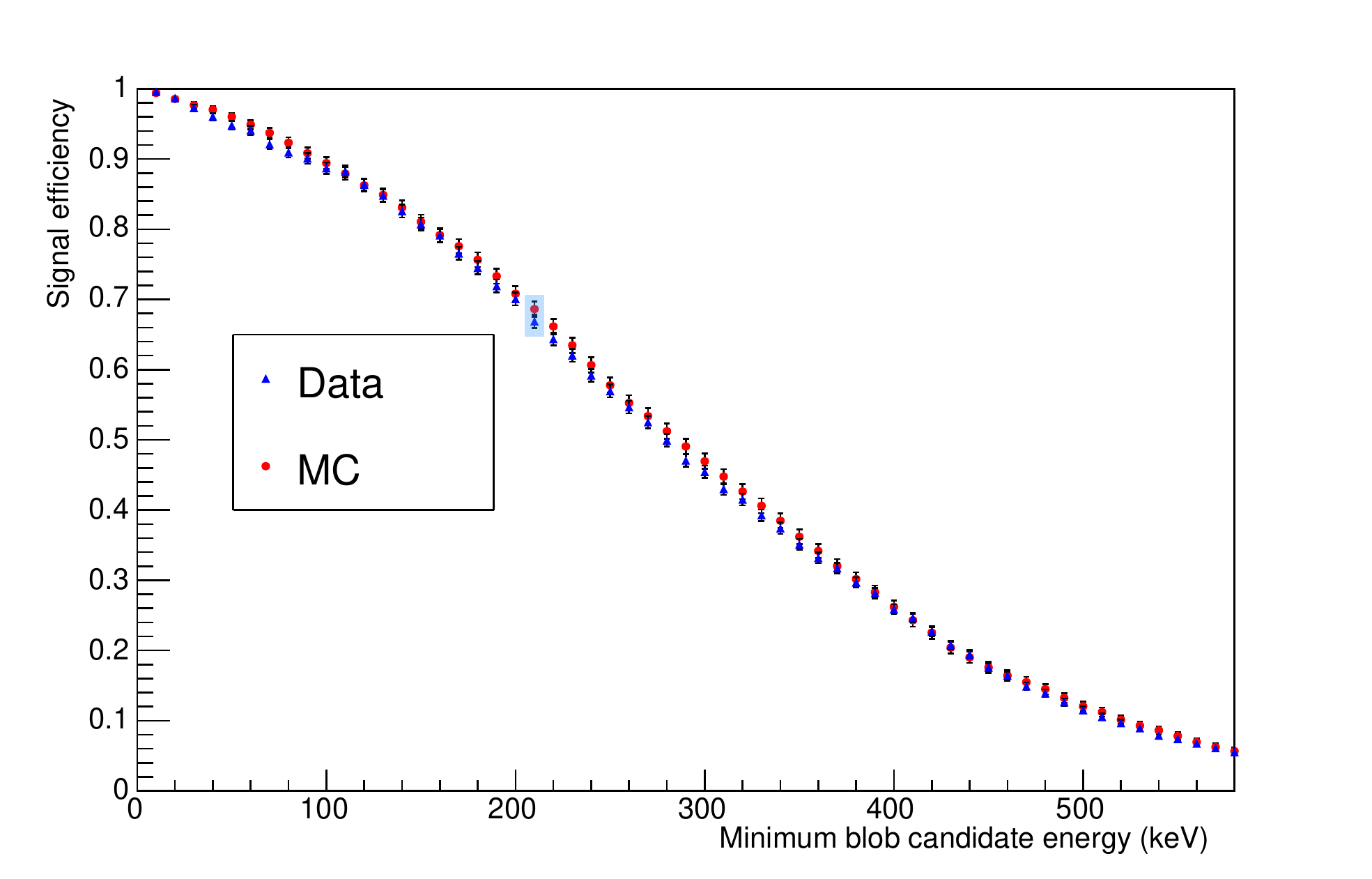} %
    \includegraphics[width=.49\textwidth]{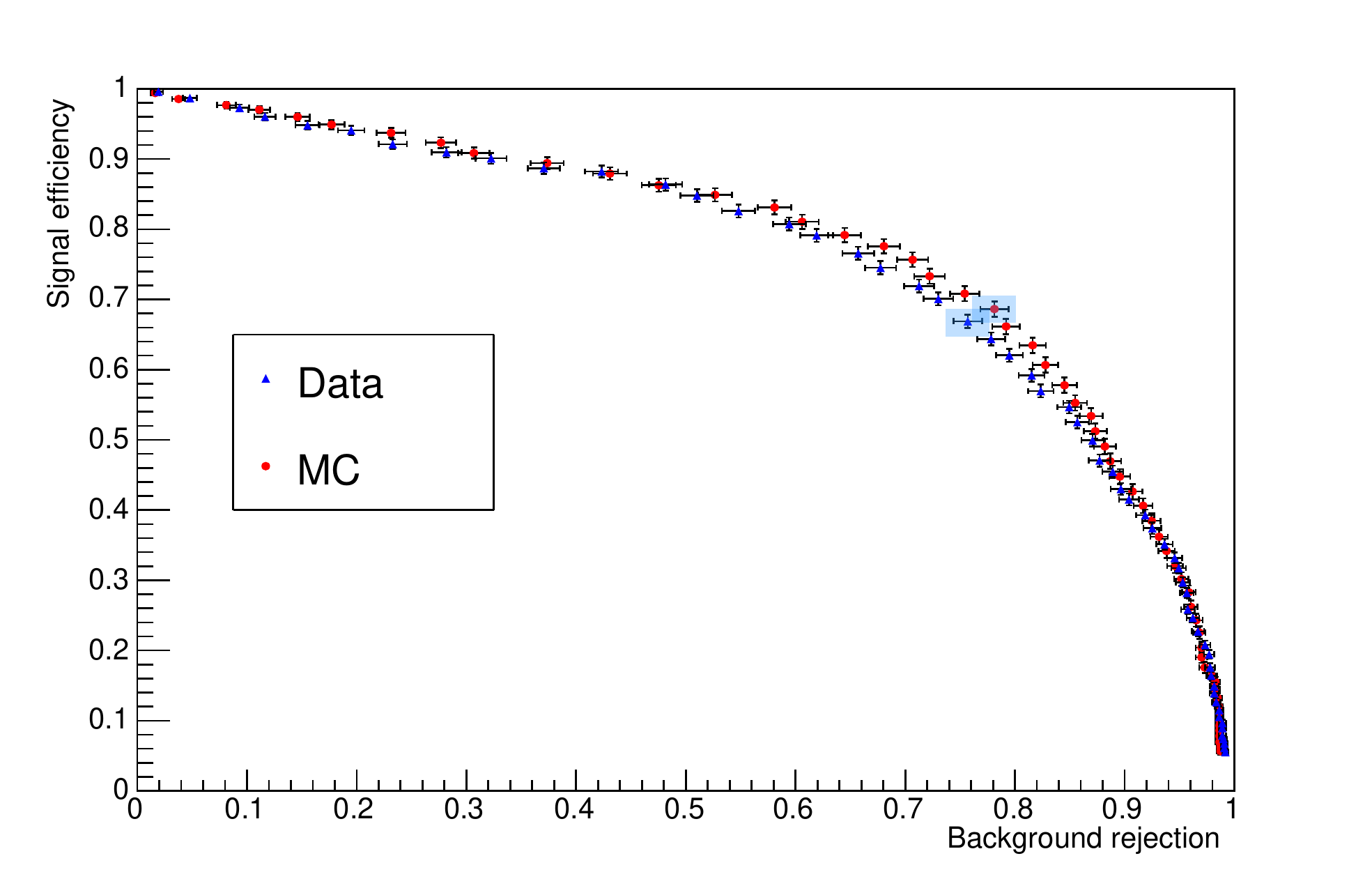} %
   \caption{\small (Left) Signal
     efficiency as a function of the required minimum energy of the lower energy blob candidate. Both data and Monte Carlo simulation are shown and the values corresponding to the cut used in this analysis are highlighted. (Right) Signal efficiency as a function of  background rejection (proportion of background events removed
     from the sample by the two-blob cut) varying the required minimum
     energy of the lower energy blob candidate. }
    \label{fig:sigVsBkg}
  \end{center}
\end{figure}

Fig.~\ref{fig:sigVsBkg}-\emph{left} shows the final two-blob selection
for both data and MC, displaying signal efficiency for the double
escape peak events and background rejection (defined as the proportion
of events that does not pass the two-blob selection) varying the
minimum energy required for the lower energy
blob. Fig.~\ref{fig:sigVsBkg}-\emph{right} shows signal efficiency as
a function of the value of the energy requirement. Data and MC
simulations are in agreement to within 2 statistical sigmas for all
points, a further validation of the NEXT Monte Carlo.


\subsection{Discussion}
\label{subSec:Discussion}
The power of the topological signature in rejection of background and
acceptance of signal events in a high-pressure xenon gas TPC (NEXT-DEMO)
is well described by the detailed simulation and reconstruction of the
Monte Carlo events (the agreement between data and MC is better than 3\%).
The NEXT-DEMO MC samples have
also been simulated using the `fast simulation' which forms voxels
using a Gaussian smearing of the energies of the true deposits. The
efficiencies obtained for the two-blob cut are in agreement within 5\%
with the results of this paper. This agreement allows for an informed
comparison between these results and those of
Ref. \cite{bckgrmodel} where the fast simulation of NEXT-100 was used
to analyse events from \bbonu, \BI, and \TL.

Compared to the present analysis, the fast simulation and analysis of NEXT-100 predict significantly improved background survival rate of 10\% as opposed to 24\% for the
NEXT-DEMO configuration, at the same signal efficiency. A major difference which affects these results is
the relative size of the two detectors. NEXT-100 has a drift length of
1.3~m with a circular cross section of $\sim$1~m and will operate at
15~bar pressure making it easily large enough to contain electron
tracks at energies similar to \Qbb\ regardless of topology or
orientation. NEXT-DEMO, on the other hand, is significantly smaller
resulting in the fiducial cuts favouring more tortuous tracks which do
not displace as far from their origin and are more difficult to
reconstruct and more prone to blob candidate overlap. This bias, which
will not be present in NEXT-100, is expected to account for the
differences observed.


The topological signatures and background rejection capabilities of
the algorithm presented in this paper demonstrate the adequate
performance for the proposed NEXT-100 experiment. Future
experiments may require larger, one tonne scale, detectors and
also better background rejection. The high potential of the
topological signatures has been demonstrated by the Gotthard 
experiment, where only 3.5\% of the background events were allowed to 
pass the final selection requirements, albeit with the detector
configuration which compromised the energy resolution \cite{Luscher:1998sd}. It is expected
that future improvements in the design of the high-pressure xenon gas
TPC and more accurate reconstruction algorithms will provide
significant improvements above the level expected for NEXT-100 detector.

%% file: src/Conclus.tex
The possibility to identify single and double electron topologies in a xenon gas TPC has been demonstrated using the NEXT-DEMO detector. This topological separation provides a powerful handle for the rejection of the mainly single electron backgrounds faced by gaseous xenon \bbonu\ experiments. In the NEXT-100 experiment a reduction of an order of magnitude in background is expected using this method.

Modelling signal using topologies left by electron-positron pairs induced by gammas from \TL\ decay and background using single electrons induced by the de-excitation gamma produced as part of the \NA\ chain, the topological separation has been optimised and applied to both Monte Carlo and data. Considering only the final requirement where a minimum end-point energy of 210~keV was required, a signal efficiency of $66.7 \pm 0.9$ (stat.) $\pm 0.3$ (fit)\% was measured. The same selection results in $24.3 \pm 1.4$ (stat.)\% of background events entering the final sample. A parallel study using the simulation of \bbonu\ events and backgrounds at that energy in NEXT-100 has found a similar efficiency, for roughly half of the background \cite{bckgrmodel}. This difference can be accounted for considering the difference in size both of the events themselves and of the detectors. Higher energy events tend to be longer, making it easier to differentiate MIP-like sections from the end-points of the events improving the rejection of backgrounds at \Qbb. The small radius of NEXT-DEMO coupled with the requirement of full containment also favours less extended events where the reconstruction is more complicated. The analysis will be repeated using data from the next stage of the NEXT experiment which uses radiopure materials in its construction and has a larger fiducial volume.